\documentclass[a4paper,twoside]{article}
\usepackage{amssymb}
\usepackage{amsmath,mathrsfs}
\usepackage{amsfonts}
\usepackage{graphicx}
\usepackage{graphicx}
\begin{document}

\title{\Large{\bf 
Loop quantum gravity and black hole singularity 
}}
\author{\\ Leonardo Modesto
 \\[1mm]
\small{ Department of Physics, Bologna University V. Irnerio 46, I-40126 Bologna \& INFN Bologna, EU}
   }

\date{\ } 
\maketitle

\begin{abstract}
In this paper we summarize {\it loop quantum gravity} (LQG)
 and we show how ideas developed in LQG can solve the black hole singularity 
problem when applied to a minisuperspace model. 
\end{abstract}

\section*{Introduction}
Quantum gravity is the theory by which we try to reconcile general relativity and quantum
mechanics. 
Because in general relativity the space-time is dynamical,  
it is not possible to study other interactions on a fixed background. 
 The theory called
 ``loop quantum gravity" (LQG) \cite{book}
is the most widespread nowadays and it is 
one of the non perturbative and 
background independent approaches to quantum gravity 
(another non perturbative approach to quantum gravity is called  {\em asymptotic safety 
quantum gravity} \cite{MR}). 
LQG is a quantum geometrical 
fundamental theory that reconciles general relativity 
and quantum mechanics at the Planck scale.
The main problem nowadays is to connect this fundamental theory
with standard model of particle physics and in particular with the 
effective quantum field theory.
In the last two years great progresses has been done to
connect LQG with the low energy physics by
the general boundary approach \cite{MV}, \cite{ModestoRovelli}.
Using this formalism it has been possible 
to calculate the graviton propagator in four \cite{Rovelli1}, \cite{BMSR}
and three dimensions \cite{Simone12}. 
In three dimensions it has been showed that a noncommuative 
field theory can be obtained from spinfoam models 
\cite{Freidel}. 
Similar efforts in four dimension are going in progress \cite{FB}.
{\em Algebraic quantum gravity}, a theory inspired by LQG, 
contains quantum field theory on curved space-time as low energy limit
\cite{giesel}.
About an unified theory of particle physics and gravity
authors in a recent paper \cite{braids} have showed that  
spinfoam models \cite{book}, including loop quantum gravity,
are also unified theories, in which matter degrees of freedom are automatically included
and a complete classification of the {\em standard model} spectrum
is realized.

Early universe and black holes 
are other interesting places for testing the validity of LQG.
In the past years 
applications of LQG ideas to minisuperspace models 
lead to some interesting results in those fields. 
In particular it has been 
showed in cosmology \cite{Boj}, \cite{MAT} and recently in black hole physics
\cite{work1}, \cite{work2}, \cite{work3}, \cite{ABM} that it is 
possible to solve the cosmological singularity problem and the
black hole singularity problem by using 
tools and ideas developed in full LQG. 
Other recent results concern a semiclassical analysis of the black
hole interior \cite{BHI} and the evaporation process \cite{ELQBH}.

We can summarize the ``loop quantum gravity
program" in the following research lines
\begin{itemize}
\item the first one dedicated to obtain 
quantum field theory from the fundamental theory
rigorously defined;
\item the second one
dedicated to apply LQG to cosmology and 
black holes where extreme energy conditions need to
know a quantum gravity theory.
\end{itemize}

The paper is organized as follow.  In the first section we briefly
recall loop quantum gravity at kinematical and 
dynamical level.
In the second section we recall a simplified model \cite{work1}
showing how quantum gravity solves the black hole singularity 
problem. In the third section we 
summarize ``loop quantum black hole" (LQBH) \cite{ABM}.
This is a minisuperspace model 
inspired by LQG 
where we quantize   
the Kantowski-Sachs space-time
without approximations. This model is useful to understand the 
black hole physics near the singularity because the space-time 
inside the event horizon is of Kantowski-Sachs type.

\section{Loop quantum gravity in a nutshell}

In this section we recall the structure of the theory introducing the Ashtekar's formulation 
of general relativity \cite{variables}, the kinematical Hilbert space,quantum geometry and   
quantum dynamics.
\subsection{Canonical gravity in Ashtekar variables}
The classical  starting point of LQG \cite{book} is the Hamiltonian formulation of general relativity. 
In ADM Hamiltonian 
formulation of the Einstein theory, the fundamental variables are the three-metric 
$q_{ab}$ of the spatial section $\Sigma$ of a foliation of the four-dimensional manifold 
$\mathcal{M} \cong \mathbb{R} \times \Sigma$, and the extrinsic curvature $K_{ab}$.
In LQG the fundamental variables are the Ashtekar variables: they consist  
on an $SU(2)$ connection $A_a^i$ and the electric field $E_i^a$, where $a, b, c, \dots = 1, 2, 3$ are tensorial indices on the spatial section and $i, j, k, \dots = 1, 2, 3$ are indices in the 
$su(2)$ algebra.   
The density weighted triad $E^a_i$ is related to the triad $e^i_a$ by the 
relation $E^a_i = \frac{1}{2} \epsilon^{abc} \, \epsilon_{ijk} \, e^j_b \, e^k_c$.  The 
metric is related to the triad by $q_{a b} = e^i_a \, e^j_b \, \delta_{ij}$.
Equivalently, 
\begin{equation}
\sqrt{\mbox{det}(q)} \, q^{a b} = E^a_i \, E^b_j \, \delta^{ij}.    
\end{equation}
The rest of the relation between the variables $(A^i_a, E^a_i)$ and the ADM variables 
$(q_{ab}, K_{ab})$ is given by
\begin{eqnarray}
A^i_a = \Gamma^i_a + \gamma \, K_{ab} E^b_j \delta^{i j}
\end{eqnarray} 
where $\gamma$ is the Immirzi parameter and $\Gamma^i_a$ is the spin 
connection of the triad, namely the solution of Cartan's equation: 
$\partial_{[a} e^i_{b]} + \epsilon^i_{j k} \, \Gamma^j_{[a} e^k_{b]} = 0$.

The action is 
\begin{eqnarray}
S = \frac{1}{\kappa \,\gamma} \int dt \int_{\Sigma} d^3 x 
\left[- 2 \mbox{Tr}(E^a \dot{A}_{a}) - N \mathcal{H} - N^a \mathcal{H}_a - N^i \mathcal{G}_i \right],
\label{action}
\end{eqnarray}  
where $N^a$ is the shift vector, $N$ is the lapse function and $N^i$ is the Lagrange
multiplier for the Gauss constraint $\mathcal{G}_i$. 
We have introduced also the notation $E = E^a \partial_a = E^a_i \tau^i \partial_a$
and $A = A_{ a} d x^a = A^i_a \tau^i d x^a$.
The functions $\mathcal{H}$, 
$\mathcal{H}_a$ and $\mathcal{G}_i$
are respectively the Hamiltonian, diffeomorphism and Gauss constraints, given by
\begin{eqnarray}
&& \mathcal{H}(E^a_i, A^i_a) = 
- 4 \, e^{-1} \, \mbox{Tr} \left(F_{ab} \, E^a E^b \right) 
- 2 \, e^{-1} \, (1+ \gamma^2) \, E^a_i E^b_j K^i_{[a} K^j_{b]}
\nonumber \\
&& \mathcal{H}_b(E^a_i, A^i_a) = E^a_j \, F^j_{ab} - (1+ \gamma^2) K^i_a G_i \nonumber \\
&& \mathcal{G}_i(E^a_i, A^i_a) = \partial_a E^a_i + \epsilon_{ij}^k \, A^j_a E^a_k, 
\label{constraints}
\end{eqnarray}
where the curvature field strength is 
$F_{ab} = \partial_a A_{ b} - \partial_b A_{a} + \left[A_{ a}, A_{ b} \right]$ and
$e = \mbox{det}(e^i_a)$. 
The constraints (\ref{constraints}) are respectively generators for the
foliation reparametrization, for the $\Sigma$ surface reparametrization and for
the gauge transformations.
   The symplectic structure for the Ashtekar Hamiltonian formulation of general 
   relativity is  
\begin{eqnarray}
\left\{E^{a}_j(x),A_{b}^i(y)\right\}=\kappa \, \gamma
\delta^a_{b}\delta^i_{j} \delta(x,y), \ \ \ \
\left\{E^{a}_j(x),E^{b}_i(y)\right\}=\left\{A_{a}^j(x),A_{b}^i(y)\right\}=0.
\end{eqnarray}
General relativity in metric formulation is defined by the 
Einstein equations $G_{\mu \nu} = 8 \pi G_N \, T_{\mu \nu}$.
The Ashtekar Hamiltonian formulation of general relativity is instead 
defined by the constraints $H=0$, $H_a=0$, $G_i=0$ and by the Hamilton
equations of motion: $\dot{A}^i_a = \{A^i_a, \mathcal{H} \}$ and $ \dot{E}^a_i = \{E^a_i, \mathcal{H} \}$.

\subsection{The Dirac program for quantum gravity}
The general strategy to quantize a system with constraints was introduced by Dirac. 
The program consist on :
\begin{enumerate}
\item find a representation of the phase space variables of the theory
as operators
in an auxiliary kinematical Hilbert space $H_{kin}$ satisfying the
standard commutation relations, i.e., $\{\ ,\ \}\rightarrow
-i/\hbar [\ ,\ ]$;
\item promote the constraints to (self-adjoint) operators in $H_{kin}$. 
For gravity we must quantize a set of seven constraints $\mathcal{G}_i(A,E)$,
$\mathcal{H}_a(A,E)$, and $\mathcal{H}(A,E)$ and we must solve the
quantum Einstein's equations (for $\gamma = i$)
\begin{eqnarray}
&& \hat{\mathcal{H}} | \psi \rangle =  
\left[- 4 \, e^{-1} \, \mbox{Tr} \left(F_{ab} \, E^a E^b \right) + \mathcal{H}_{M} \right] | \psi \rangle = 0,
\nonumber \\
&& \hat{\mathcal{H}}_b | \psi \rangle = \left[E^a_j \, F^j_{ab} + \mathcal{H}_{M b} \right]| \psi \rangle = 0, 
\nonumber \\
&& \hat{\mathcal{G}}_i |\psi \rangle = \left[\partial_a E^a_i + \epsilon_{ij}^k \, A^j_a E^a_k + \mathcal{G}_{M i} \right] | \psi \rangle = 0
\label{Quanconstraints}
\end{eqnarray}
We will consider pure gravity then the matter constraints are identically zero. 
\item introduce an inner product defining the physical Hilbert space $H_{phys}$.
\end{enumerate}

\subsection{Kinematical Hilbert space}
The fundamental ingredient of LQG is the holonomy of the Ashtekar connection
along a path $e$,  
$h_{e}[A] = P \exp -\int_{e} A \in SU(2)$.   
Given two oriented paths $e_1$ and $e_2$ such that
the end point of $e_1$ coincides with the starting point of $e_2$
so that we can define $e=e_1e_2$ we have the composition rule  
$h_{e}[A] = h_{e_1}[A] h_{e_2}[A]$.
By a gauge transformation the holonomy transforms as 
\begin{eqnarray} 
h^{\prime}_{e}[A]=g(x(0))\ h_{e}[A]\ g^{-1}(x(1)),
\label{gauge}
\end{eqnarray}
 and by a 
Diffeomorfism of the three dimensional variety  
 $ \phi \in {\rm Diff}(\Sigma)$ we have 
\begin{eqnarray}
 h_{e}[\phi^*A]=h_{\phi^{-1}(e)}[A], 
 \label{Diff}
\end{eqnarray}
where $\phi^*A$ denotes the action of $\phi$ on the connection. 
In other words, transforming the connection with a diffeomorphism is
equivalent to simply moving the path with $\phi^{-1}$.

We introduce now the space of cylindrical functions (Cyl$_\gamma$)
where $\gamma$ denotes a general graph.
A graph $\gamma$ is defined as a collection of paths $e \subset \Sigma$
 ($e$ stands for edge) meeting at most at their endpoints. 
If $N_e$ is the number of paths or edges of the graph
and 
$e_i$, for $i=1,\cdots N_e$, are the edges
of the corresponding graph $\gamma$ a
cylindrical function is an application $f: SU(2)^{N_e}\rightarrow \mathbb{C}$, 
defined by 
\begin{eqnarray} 
\psi_{\gamma,f}[A]:=f(h_{e_1}[A],h_{e_2}[A],\cdots
h_{e_{N_e}}[A]).
\label{Cylpsi}
\end{eqnarray}
Two particular examples of cylindrical functions are 
the holonomy around a loop, $W_{\gamma}[A]:={\rm Tr}[h_{\gamma}[A]]$
and the three edges function  
$\Theta^{1,1/2,1/2}_{e_1 \cup e_2 \cup
e_3}[A]= \,^{1}{\mathscr D}(h_{e_1}[A])^{ij} \, 
^{1/2} {\mathscr D}(h_{e_2}[A])_{AB} \, 
^{1/2} {\mathscr D} (h_{e_3}[A])_{CD} \, 
f^{ABCD}_{ij}$,
where $^j{\mathscr D}(h_{e_i})$ is the $SU(2)$ representation 
for the holonomy along the path $e_i$ and $f^{ABCD}_{ij}$ are complex coefficients.
The algebra of generalized connections is given by 
${\rm Cyl}=\cup_{\gamma} {\rm Cyl}_{\gamma}$.

We introduce now the space of 
{\em spin networks} states. 
We label the set of edges
$e\subset \gamma$ with spins $\{j_e\}$. To each node $n\subset
\gamma$ one assigns an invariant tensor, called {\em intertwiner},
$\iota_n$, in the tensor product of representations labelling the
edges converging at the corresponding node. 
The spin network function is defined
\begin{equation}
s_{\gamma,
\{j_{e}\},\{\iota_{n}\}}[A]=\bigotimes_{n \subset \gamma} \
\iota_n\ \bigotimes_{e \subset \gamma}\
\,^{j_{e}} {\mathscr D}(h_{e}[A]) \;,
\label{SpinNet}
\end{equation}
where the indices of representation matrices and invariant tensors
are implicit to simplify the notation. 
An example of spin network state is 
\begin{eqnarray}
\Theta^{1,1/2,1/2}_{e_1 \cup e_2 \cup
e_3}[A]= \,^{1} {\mathscr D}(h_{e_1}[A])^{ij} \, 
^{1/2} {\mathscr D}(h_{e_2}[A])_{AB} \, 
^{1/2} {\mathscr D} (h_{e_3}[A])_{CD} \, 
\sigma_{i}^{AC} \, \sigma_{j}^{BD},
\end{eqnarray}
where for the particular representations converging to the two three-valent nodes
of the graph the intertwiner tensor is the Pauli matrix. The spin network states 
are gauge invariant because for any node of the graph we have invariant 
tensors (intertwiners), then on the spin network states the Gauss constraint is solved 
as asked from the Dirac program of the previous subsection.

To complete the Hilbert space definition we must introduce an inner product.
The scalar product is defined by the 
Ashtekar-Lewandowski measure 
\begin{eqnarray}
<\psi_{\gamma,f},\psi_{\gamma^{\prime},g}> =
\mu_{AL}(\overline{\psi_{\gamma,f}}\psi_{\gamma^{\prime},g})=
\int \prod\limits_{e\subset \Gamma_{\gamma\gamma^{\prime}}}
dh_e\ \overline{f(h_{e_1},\dots h_{e_{N_e}})} g(h_{e_1},\dots
h_{e_{N_e}}),
\label{SP}
\end{eqnarray}
where we use Dirac notation and 
$f(h_{e_1},\dots h_{e_{N_e}}), \, g(h_{e_1},\dots h_{e_{N_e}})$
are cylindrical functions; 
$\Gamma_{\gamma\gamma^{\prime}}$ is any graph
such that both $\gamma \subset \Gamma_{\gamma\gamma^{\prime}}$ and
$\gamma^{\prime} \subset \Gamma_{\gamma\gamma^{\prime}}$;
$d h_e$ is the Haar measure of $SU(2)$. 
The scalar product (\ref{SP}) is non zero only if the two cylindrical 
functions have support on the same graph. 
The kinematical Hilbert space $H_{kin}$ is the Cauchy completion of the
space of cylindrical functions Cyl in the Ashtekar-Lewandowski measure. 
In other words, in addition to
cylindrical functions we add to $H_{kin}$ the limits of all the Cauchy
convergent sequences in the norm defined by the inner product :
$\psi = \sum_{n=1}^{\infty} a_n \, \psi_n$, $||\psi||^2 = \sum_{n=1}^{\infty} |a_n|^2 \, ||\psi_n ||^2$.

We complete the construction of the theory at 
kinematical level solving the diffeomorphism constraint. 
Given $\psi_{\gamma,f}\in {\rm Cyl}$
the finite action of a Diff. transformation is implemented by
an unitary operator ${\cal U_D}$ such that 
\begin{eqnarray}
{\cal U_{D}}[\phi] \psi_{\gamma,f}[A] = \psi_{\phi^{-1}\gamma,f}[A].
\end{eqnarray}
The states invariant under Diff. transformations satisfy  
${\cal U_{D}}[\phi]\psi=\psi$ and are distributional states in the dual space
of $H_{kin}$,
$\psi \in {\rm Cyl}^{\star}$
\begin{eqnarray}
([\psi_{\gamma,f}]|=\sum_{\phi \in {\rm Diff}(\Sigma)}
<\psi_{\gamma,f}|{\cal U_{D}}[\phi]= \sum_{\phi \in {\rm Diff}(\Sigma)}
<\psi_{\phi\gamma,f}|,
\end{eqnarray}
where the sum is over all diffeomorphisms which modified the spin network.
The brackets in $([\psi_{\gamma,f}]|$ denote that the
distributional state depends only on the equivalence class
$[\psi_{\gamma, f}]$ under diffeomorphisms. Clearly we have
$([\psi_{\gamma,f}]|{\cal U_{D}}[\alpha]=([\psi_{\gamma,f}]| \,\,   \forall \,\, \alpha \in {\rm Diff}(\Sigma)$.

We conclude that the Dirac's program at kinematical level is realized by the
{\em Gelfand triple} 
\begin{eqnarray}
{\rm Cyl} \subset H_{kin} \subset {\rm Cyl^*} \,\, \stackrel{SU(2)}{\rightarrow} \,\, {\rm Cyl_0} \subset H_{kin} \subset {\rm Cyl^*_0} \,\, \stackrel{{\rm Diff.}}{\rightarrow} \,\, H_{\rm Diff} \subset {\rm Cyl}^*,
\end{eqnarray}
where ${\rm Cyl_0}$ is the subspace of cylindrical functions invariant under $SU(2)$.

At quantum level the phase space variables operators are represented 
on the spin network space 
by the holonomy operator $\hat{h}_{e}[A]$ that acts multiplicatively 
on the states 
and  by the smearing of the triad $E^a_i$ on a two dimensional surface $S \in \Sigma$
\begin{eqnarray}
 \hat{E}[S,\alpha]=\int \limits_S
d\sigma^1d\sigma^2\frac{\partial x^{a}}{\partial \sigma^{1}}
\frac{\partial x^{b}}{\partial \sigma^{2}} \alpha^i
\hat{E^a_i}\epsilon_{abc}=-i\hbar \kappa \gamma\int\limits_S
d\sigma^1d\sigma^2\frac{\partial x^{a}}{\partial \sigma^{1}}
\frac{\partial x^{b}}{\partial \sigma^{2}}\alpha^i
\frac{\delta}{\delta A_c^i} \epsilon_{abc}, 
\label{pippo}
\end{eqnarray}
where $ \alpha^i$ is the smearing function with values on the Lie algebra of $SU(2)$.
The action of $\hat{E}[S,\alpha]$ on the spin network states can be calculated using  
\begin{eqnarray}
&& \hat{E}[S,\alpha]h_e(A)=-i l_P^2\gamma
\alpha^i h_{e_1}(A)\tau_i h_{e_2}(A) \, ,   \,\,\,\, e = e_1 \cup e_2, \nonumber \\
&& \hat{E}^a_i(x) h_e(A) = -i l_P^2 \frac{\delta}{\delta A^i_a} h_e(A) =
\int ds \, \dot{e}^a(s) \delta^3(e(s),x)h_{e_1}(A) \tau_i h_{e_2}(A),
\label{Eact}
\end{eqnarray}
where $\dot{e}^a(s)$ is tangent to the curve $e(s)$ in the graph $\gamma$.
The pair $(\hat{h}_{e}[A], \hat{E}[S,\alpha] )$ realizes the first point of the 
Dirac's program. 

\subsection{Quantum geometry and dynamics}
\begin{figure}
 \begin{center}
  \includegraphics[height=3.5cm]{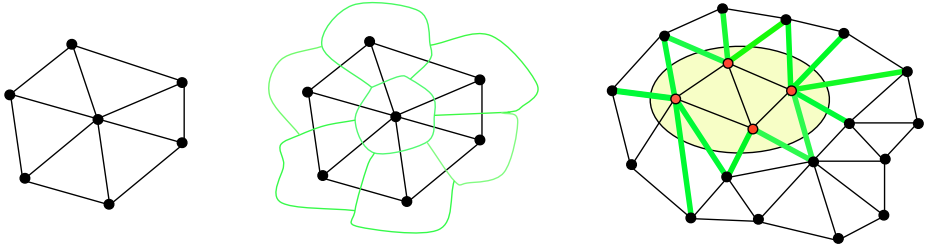}
  \end{center}
  \caption{\label{Chunk} 
                 In this picture we show the spin network physical meaning.
                 The graph on the left represents a particular spin network. In the center we 
                 represent the same spin network and the dual decomposition of the space section
                 in {\em chunk of space}. In the last picture on the right we consider another 
                 spin network and a particular dual volume region.
                 The yellow region is a chunk of space with volume eigenvalues related 
                  to the red intertwiners and area eigenvalues given by the $SU(2)$ representations 
                  associated to the green edges.  
                   }
  \end{figure}

We are going to give a physical interpretation of the Hilbert space
previously introduced.
We consider the spatial section $\Sigma$ of the space-time 
and we study the spectrum of the area $S$ and volume $R$ 
in the section $\Sigma$ \cite{LoopOld}.  
We define the area of a surface $S$ as the limit of a Riemann sum
\begin{eqnarray}
A_S= \lim_{N\rightarrow \infty} A_S^N \,\, , \,\,\,\,\,\, A_S^N=\sum_{I=1}^N
\sqrt{E_i(S_I)E^i(S_I)} 
\end{eqnarray}
where $N$ is the number of cells. The quantum area operator is 
$\widehat{A}_S= \lim_{N\rightarrow \infty} \widehat{A}_S^N$.
Using (\ref{Eact}) we calculate the action 
$ \widehat
E_i(S_I) \widehat E^i(S_I)
\,^j{\mathscr D}(h_e[A])_{mn}=(l_P^2\gamma)^2
(j(j+1))\,^j{\mathscr D}(h_{e}[A])_{mn}$.
The area spectrum
for spin network without edges and nodes on the surface is 
\begin{eqnarray}
 \hat{A}_S|\gamma, j_e, \iota_n \rangle= l_P^2 \gamma \sum_{p\cap S} \sqrt{j_p(j_p+1)} |\gamma, j_e, \iota_n\rangle,
\end{eqnarray}
where $j_p$ are the representations on the edges 
that cross the surface $S$.
Now we consider a region $R$ with a number $n$ of nodes inside.
The spectrum of the volume operator for the region $R$ is
\begin{eqnarray}
\hat{V}_{R} |\gamma, j_e, \iota_n\rangle = l_P^3 \gamma^{3/2} 
\sum_{n \subset R} \sqrt{w^{(\iota_n)}(j_n)} |\gamma, j_e, \iota_n\rangle.
\end{eqnarray}
We have all the ingredients to give a physical 
interpretation of the Hilbert space.
The states have support on graphs that are a collection of nodes 
and edges converging in the nodes.
The dual of a spin network corresponds to a
discretization of the three dimensional surface $\Sigma$. The dual of a 
set of edges is the 2-dimensional surface crossed by the links
and the dual of a set of nodes is the volume {\em chunk} contained 
nodes (see Fig.\ref{Chunk}). 
   
We must now implement the Hamiltonian 
constraint on the Hilbert space. The Euclidean part 
of the constraint $S^E(N) = \int_{\Sigma} d^3 x N(x) \mathcal{H}(E^a_i, A^i_a)$, 
\begin{eqnarray}
 S^{E}(N)=\int\limits_{\Sigma} d^3 x \ N(x)
\frac{E_i^aE_j^b}{\sqrt{{\rm det}(E)}}\epsilon^{ij}_{\ \ k}
F_{ab}^k.
\label{EEF}
\end{eqnarray}
Using the Thiemann's trick \cite{book} we can express the inverse of $\sqrt{{\rm det}(E)}$ by 
\begin{eqnarray}
\frac{E_i^bE_j^c}{\sqrt{{\rm det}(E)}}\epsilon^{ijk}\epsilon_{abc}=\frac{4}{\kappa\gamma}\left\{A^k_a,V\right\},
\end{eqnarray}
and the Hamiltonian constraint is 
\begin{eqnarray}
S^{E}(N)= \frac{4}{\kappa \gamma} \int\limits_{\Sigma} dx^3 \ N(x) \ \epsilon^{abc}
\delta_{ij} F^i_{ab}\left\{ A^j_c,V \right\}.
\end{eqnarray}
Now we define the Hamiltonian constraint in terms of holonomies.
Given an infinitesimal loop $\alpha_{ab}$ on the $ab$-plane in the surface $\Sigma$
with coordinate area $\epsilon^2$, we can define $F^i_{ab}$ in terms of holonomies by
$h_{\alpha_{ab}}[A]-h^{-1}_{\alpha_{ab}}[A]=\epsilon^2
F_{ab}^i\tau_i+{\cal O}(\epsilon^4)$ and 
$h_{e_a}^{-1}[A]\left\{h_{e_a}[A],V\right\}=\epsilon
\left\{A^i_a,V\right\}+{\cal O}(\epsilon^2)$
($e_a$ is a path along the $a$-coordinate of coordinate length $\epsilon$). 
With these ingredients 
the quantum constraint can formally be written
\begin{eqnarray}
S^{E}(N)=\frac{4}{\kappa \gamma} \, \lim_{\epsilon\rightarrow 0}
\sum_I \ N_I \ \epsilon^{abc}{\rm Tr}\left[( \widehat
h_{\alpha^I_{ab}}[A]-\widehat h^{-1}_{\alpha^I_{ab}}[A]) \widehat
h_{e^I_c}^{-1}[A]\left[\widehat h_{e^I_c}[A],\widehat
V\right]\right],
\label{QH}
\end{eqnarray}
where we have replaced the integral by a Riemann sum over cells of coordinate volume
$\epsilon^3$.
It is easy to see that the regularized quantum
scalar constraint acts only on spin network nodes,
because in (\ref{EEF}) $F_{ab}$ and $E^a E^b$ 
are respectively antisymmetric 
and symmetric in indexes on spin network states.
In fact $E^a E^b \psi_{\gamma, f} \sim \dot{e}^a \dot{e}^b \psi_{\gamma^{\prime}, f^{\prime}}$
(this is a consequence of (\ref{Eact})).
The action of (\ref{QH}) on spin network states is 
$\widehat S_{\epsilon}(N)\psi_{\gamma,f}=\sum_{n \gamma} N_n
\widehat S^n_{\epsilon}\ \psi_{\gamma,f}$, 
where $\widehat S^{n}_{\epsilon}$ acts only on the node $n\subset \gamma$
and $N_n$ is the value of the lapse $N(x)$ at the node.
The scalar constraint modifies spin networks by creating new exceptional links
around nodes. The Euclidean constraint action on 4-valent nodes is \cite{book}
\begin{eqnarray}  
&& \hspace{-0.7cm}
\widehat S^{n}_{\epsilon} \left| \begin{array}{c}
 \includegraphics[width=1cm]{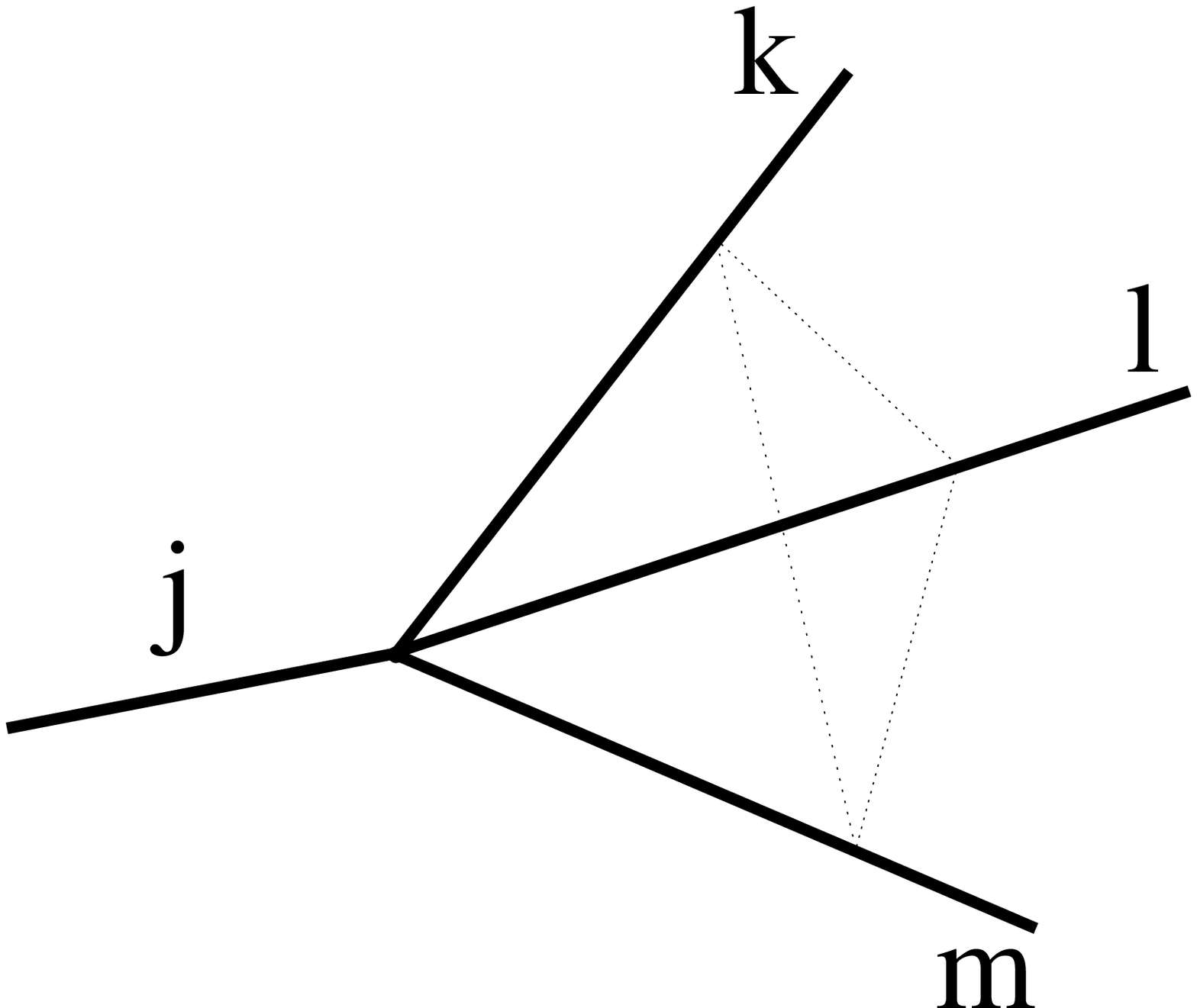} 
\end{array} \right\rangle= \sum_{op} S_{jklm,opq} \left| \begin{array}{c}
\includegraphics[width=1cm]{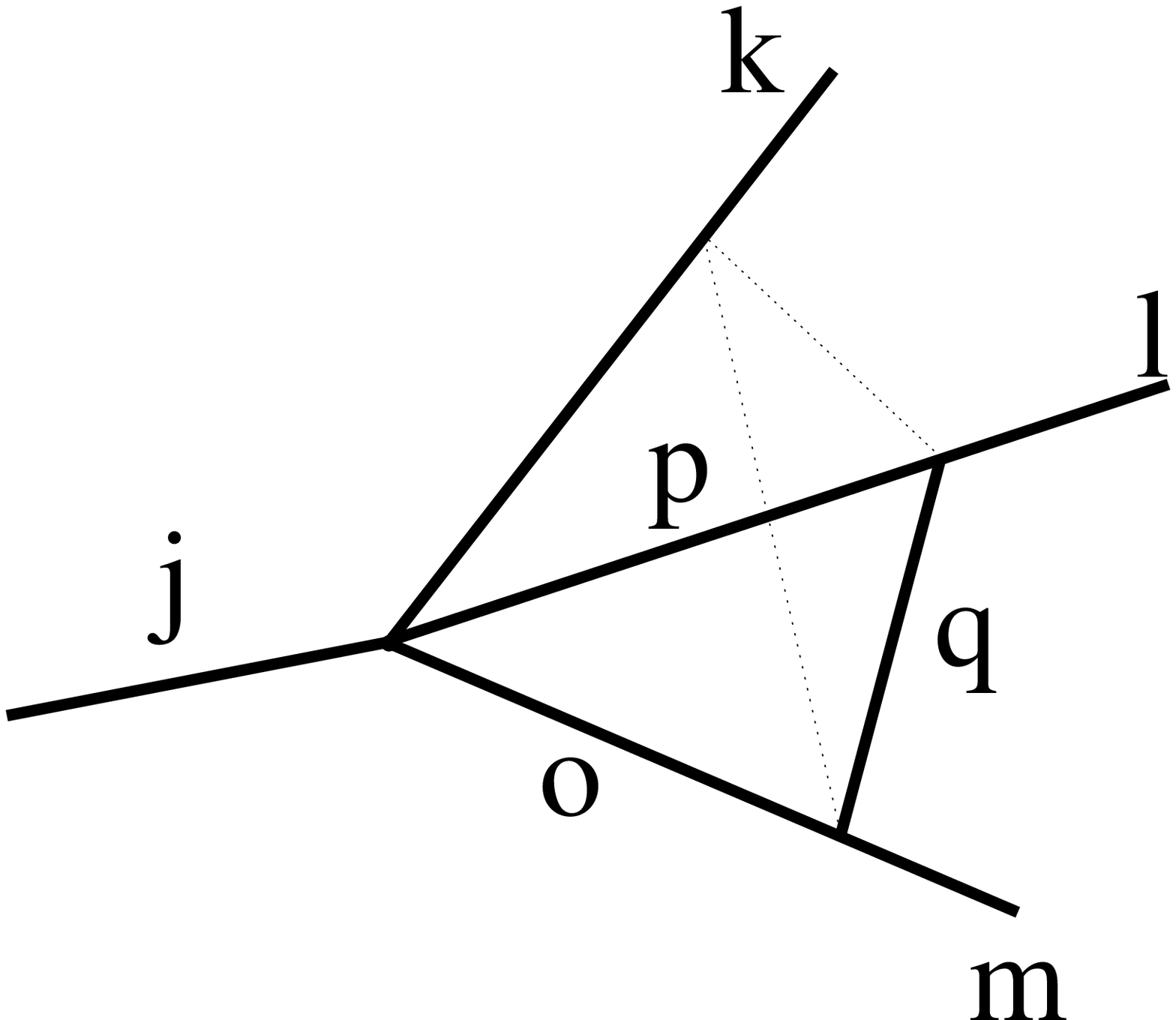}
\end{array} \right\rangle 
+ \sum_{op} S_{jlmk,opq} \left| \begin{array}{c}
\includegraphics[width=1cm]{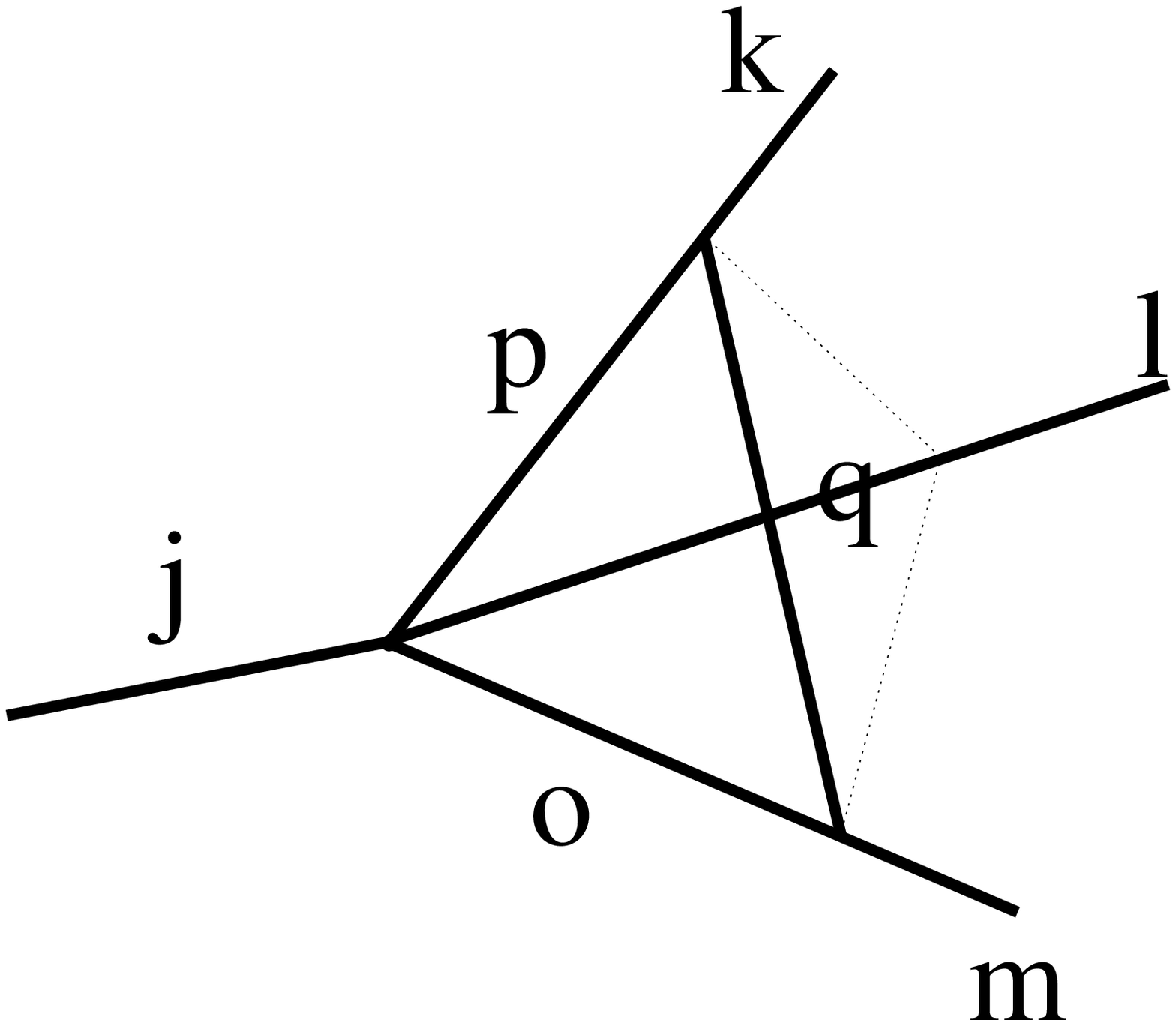}
\end{array} \right\rangle + \sum_{op} S_{jmkl,opq} \left| \begin{array}{c}
\includegraphics[width=1cm]{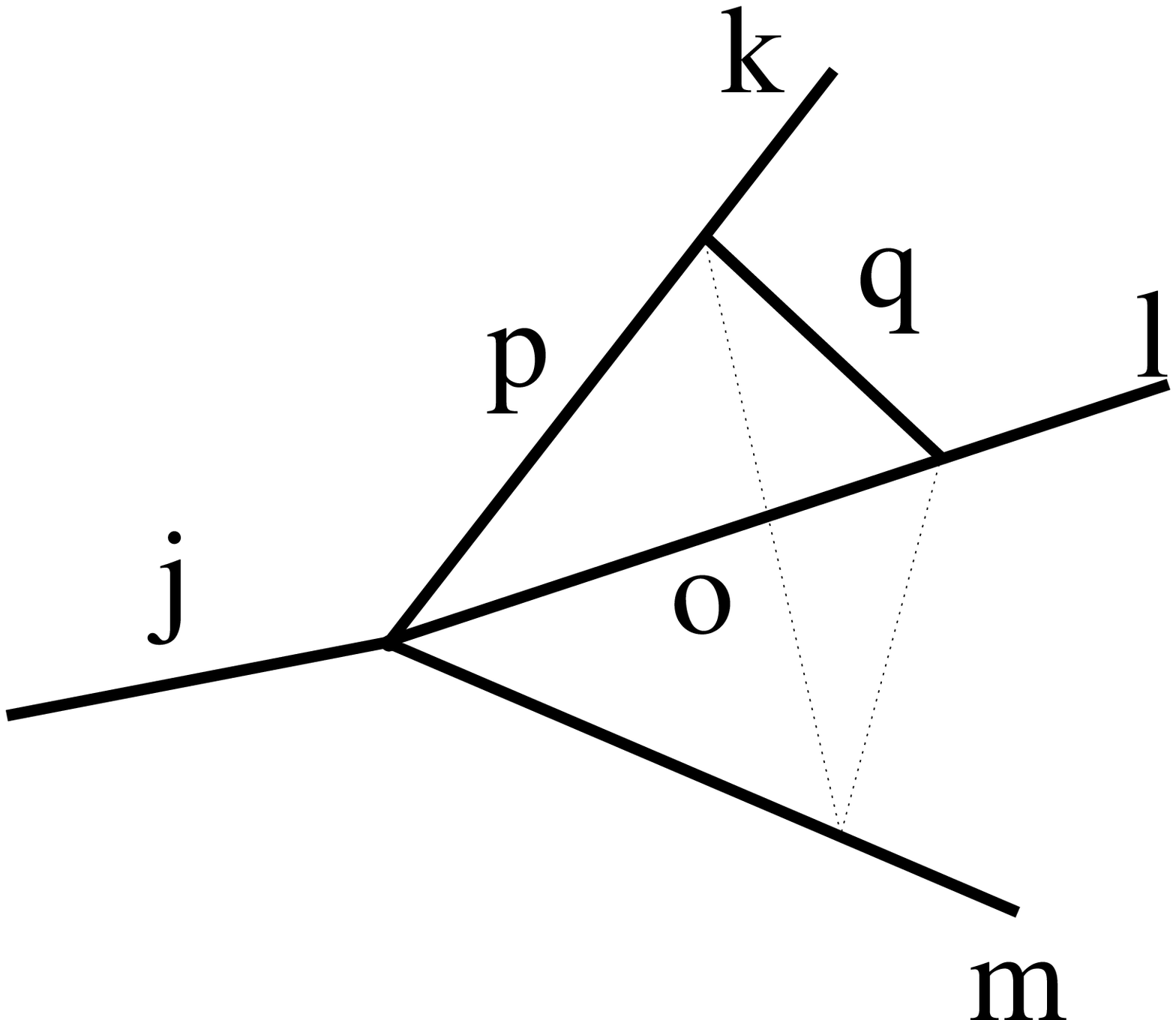} 
\end{array} \right\rangle. \nonumber \\
&&
\end{eqnarray}
The only dependence on the regularization parameter 
$\epsilon$ is in the position of the extra edges in
the resulting spin network states, then the limit $\epsilon\rightarrow 0$ 
can be defined on diffeomorphism invariant
states in $H_{Diff}$. 
The key property is that in the diffeomorphism invariant context the
position of the new link is irrelevant.
Therefore, given a diffeomorphism invariant state
$([\phi]|\in H_{Diff} \subset {\rm Cyl}^{\star}$, the quantity 
$(\phi|\widehat S_{\epsilon}(N)|\psi \rangle$
is well defined and independent of $\epsilon$. 

The operator (\ref{QH}) defines the dynamics and an unitary 
implementation of the constraint $S^E(N)$ gives the
evolution amplitude from a spin network $s$ to
a new spin network $s^{\prime}$.  
Introducing the projector operator
$P = \int D[N] \ {\rm
exp} \, (i\int  N(x) S^E(x))$
we can characterizes the solutions of {\em quantum Einstein equations}
by $P |s \rangle$, $\forall |s \rangle \in H_{kin}$.
The matrix elements of $P$ define the {\em physical scalar product}, 
$W(s, s^{\prime}) := \langle s | s^{\prime} \rangle_{phys}:= \langle s| P |s^{\prime}\rangle$.
The amplitude $W(s, s^{\prime})$ solves the Hamiltonian constraint in
the following sense. If $|\Psi_{phys} \rangle:= P | s^{\prime} \rangle$ we have that 
$S^E(N)  |\Psi_{phys} \rangle = 0$, but $\langle s |\Psi_{phys} \rangle := \Psi_{phys}(s) = W(s,s^{\prime})$, then we obtain 
\begin{eqnarray}
&& \langle s |S^E(N)  |\Psi_{phys} \rangle = \sum_{s^{\prime \prime}} 
\langle s | S^E(N)  | s^{\prime \prime} \rangle \langle s^{\prime \prime}| \Psi_{phys} \rangle= \nonumber \\
&& = \sum_{s^{\prime \prime}} S^E(N)_{s s^{\prime \prime}} \Psi_{phys}(s^{\prime \prime})=
  \sum_{s^{\prime \prime}} S^E(N)_{s s^{\prime \prime}} W(s^{\prime \prime}, s^{\prime}) = 0.
\label{W}
\end{eqnarray}
Relation (\ref{W}) shows the amplitude $W(s, s^{\prime})$ is in the 
Hamiltonian constraint kernel. $W(s, s^{\prime})$ realizes the Dirac's program
because corresponds to the finite implementation of the scalar constraint 
on the kinematical states and defines a class of models called {\em spinfoam models}
\cite{book}.

\section{Avoidance black hole singularity in quantum gravity}
In this section we study the black hole system inside the event horizon 
in ADM variables considering a simplified minisuperspace model
\cite{work1} and in particular using the fundamental ideas suggested by full LQG 
and introduced in the first section.
The simplification consist on a semiclassical
condition which reduce the phase space from four to two dimensions.
This is an approximate model but it is useful to understand the ideas
before to quantize the Kantowski-Sachs system in Ashtekar varables.

\subsection{Classical theory}

Consider the Schwarzschild solution inside the event horizon;
the metric is homogeneous and it reads 
\begin{eqnarray}
ds^2 = - \frac{dT^2}{\left(\frac{2M G_N}{T} -1 \right)}+ \left(\frac{2M G_N}{T} -1 \right) dr^2 + T^2 (\sin^2 \theta d\phi^2 + d \theta^2).
\end{eqnarray}
This metric is a particular 
representative of the Kantowski-Sachs class  \cite{KS}
\begin{eqnarray}
ds^2 = - d t^2+ a(t)^2 dr^2 +
b(t)^2 (\sin^2 \theta d\phi^2 + d \theta^2).
\label{KS}
\end{eqnarray}
Introducing (\ref{KS}) in the Einstein-Hilber action 
we obtain $S= - \frac{R}{2 \, G_N} \int dt \Big[a \, \dot{b}^2 + 2 \, \dot{a} \, \dot{b} \, b - a\Big]$
(where $R = \int dr$ is  a cut-off on the radial cell) \cite{work1}.
We introduce in the action the classical 
relation $a^2 = 2 M G_N/b(t) -1$,  
obtaining 
\begin{eqnarray}
S = \frac{R}{2 G_N} \int dt \left[\frac{\sqrt{b}}{\sqrt{2 M G_N}}
\Big(1 - \frac{b}{2 M G_N}\Big)^{-\frac{1}{2}} \, \dot{b}^2 + 
\,\frac{\sqrt{2 M G_N}}{\sqrt{b}} \Big(1 - \frac{b}{2 M G_N}\Big)^{\frac{1}{2}}
\right].
\label{Action.Mini.ridotta} 
\end{eqnarray}
The momentum conjugates to the variable $b(t)$ is 
$
p = \frac{R \, \sqrt{b}}{G_N\sqrt{2 M G_N}} \left(1 - \frac{b}{2 M
G_N} \right)^{-1/2} \, \dot{b},
$
and the Hamiltonian constraint, by Legendre transform, is
\begin{eqnarray}
H = p \, \dot{b} - L = \left( \frac{G_N \, p^2}{2R} - \frac{R}{2G_N} \right) \left[ \frac{\sqrt{2 M G_N}}{\sqrt{b}} \left(1 -
\frac{b}{2 M G_N} \right)^{1/2} \right].
\label{HamiltonianS}
 \end{eqnarray}
We introduce a further approximation.  In quantum theory, we will be
interest in the region of the scale the Planck length $l_{P}$ around
the singularity.  We assume that the Schwarzschild radius $r_{s}
= 2 M G_N$ is much larger than this scale.
In this approximation we can write
$
1 -  b/2 M G_N \sim 1
$
and $H$ becomes 
$$
H = \left( \frac{G_N \, p^2}{2R} - \frac{R}{2G_N}\right)
\frac{\sqrt{2 M G_N}}{\sqrt{b}}.
\label{Hamiltonian.approx}
$$
In the same approximation the volume of the space section is 
$V =  4 \pi R \sqrt{2 M G_N} \, b^{3/2} := l_o \, b^{3/2}$.

The canonical pair is given by $b \equiv x$ and $p$, with Poisson
brackets $\{x,p\} = 1$.  
We now assume that $x \in \mathbb{R}$. This choice it not correct classically,
because for $x =0$ we have the singularity, but it allows
us to open the possibility that the situation be different in the
quantum theory.  We introduce an algebra of classical observables and
we write the quantities of physical interest in terms of those
variables.  We are motivated by loop quantum gravity to use the
fundamental variables $x$ and
\begin{eqnarray}
U_{\delta}(p) \equiv \exp \left(\frac{8 \pi G_N \delta}{L} \, i \,  p \right) 
\label{analogO}
\end{eqnarray}
where $\delta$ is a real parameter
(see next paragraph for a rigorous mathematical definition of $\delta$) 
and $L$ fixes the unit of length. 
 The operator (\ref{analogO}) can be seen as the
analog of the holonomy variable of loop quantum gravity.

A straightforward calculation gives
\begin{eqnarray}
 && \{x , U_{\delta}(p)\} = 8 \pi G_N \frac{i \, \delta}{L} U_{\delta}(p),
 \nonumber \\
&& U_{\delta}^{-1} \{ V^n , U_{\delta} \} = l_0^n \,
U_{\delta}^{-1} \{ |x|^{\frac{3n}{2}} , U_{\delta} \} = i \,8 \pi
G_N \, l_0^n \, \frac{\delta}{L} \frac{3 n}{2}Ê\mbox{sgn}(x)
|x|^{\frac{3n}{2} -1}.
\label{Poisson.Volume}
\end{eqnarray}
These formulas allow us to express inverse powers of $x$ in terms of a
Poisson bracket between $U_{\delta}$ and the volume $V$,
 following Thiemann's trick \cite{Thie}.  
For $n = 1/3$ (\ref{Poisson.Volume}) gives
\begin{eqnarray}
\frac{\mbox{sgn}(x)}{\sqrt{|x|}} = - \frac{2 L i}{(8 \pi G_N) l_0^{1/3} \delta} \, U_{\delta}^{-1} \{ V^{\frac{1}{3} },
U_{\delta} \}. 
\label{unosux}
\end{eqnarray}
We will use this relation in quantum mechanics 
to define the physical operators. 
We are interested to the quantity $1/|x|$ because classically
this quantity diverge for $|x| \rightarrow 0$ and produce the
singularity.  We are also interested to the Hamiltonian constraint and
the dynamics and we will use (\ref{unosux}) for writing the
Hamiltonian.

\subsection{Polymer quantization}
In this section 
we recall the polymer representation \cite{AFW}
of the Weyl-Heisenberg algebra 
and 
we compare this representation with full LQG. 
The polymer representation of the Weyl-Heisenberg algebra is unitarily {\it inequivalent} 
to the Schroedi-nger representation. Now we construct the Hilbert space $H_{Poly}$. 
First of all we define a graph $\gamma$ as a finite number of points $\{x_i\}$ on the 
real line $\mathbb{R}$. We denote by Cyl$_{\gamma}$ the vector space of function $f(k)$ 
($ f : \mathbb{R} \rightarrow  \mathbb{C} $) of the type 
\begin{eqnarray}
f(k) = \sum_{j} f_j e^{ - i x_j k}
\label{genfun2}
\end{eqnarray}
where $k \in \mathbb{R}$, $x_j \in \mathbb{R}$ and $f_j \in \mathbb{C}$ and 
$j$ runs over a finite number of integer (labelling the points of the graph). We will call
cylindrical  with respect to the graph $\gamma$ the function $f(k)$ in Cyl$_{\gamma}$.
The real number $k$ is the analog of the connections in loop quantum gravity and 
the plane wave $\mbox{e}^{-i k x_j}$ can be thought as the holonomy of the connection
$k$ along the graph $\{x_j\}$.

Now we consider all the possible graphs (the points and their number can vary 
from a graph to another) and we denote Cyl the infinite dimensional 
vector space of functions cylindrical with respect to some graph: $\mbox{Cyl}=\bigcup_{\gamma}\mbox{Cyl}_{\gamma}$. A basis in Cyl is 
given by
$\mbox{e}^{-i k x_j}$ with $\langle \mbox{e}^{-i k x_i} | \mbox{e}^{-i k x_j} \rangle = \delta_{x_i, x_j}$. $H_{Poly}$ is the Cauchy completion of Cyl or more succinctly $H_{Poly} = L_2(\bar{\mathbb{R}}_{Bohr}, d \mu_0)$, where $\bar{\mathbb{R}}_{Bohr}$ is the 
Bohr-compactification of $\mathbb{R}$ and $d\mu_0$ is the Haar measure on $\bar{\mathbb{R}}_{Bohr}$.

The Weyl-Heisenberg algebra is represented on $H_{Poly}$ by the two unitary operators 
\begin{eqnarray}
 \hat{V}(\lambda) f(k) = f(k - \lambda) \, , \,\,\,\,\, 
 \hat{U}(\delta) f(k) = \mbox{e}^{i \delta k} \, f(k),
\end{eqnarray}
where $\lambda, \delta \in \mathbb{R}$.
In terms of eigenkets of $\hat{V}(\lambda)$ (we associate a ket $|x_j \rangle$ with 
the basis elements $\mbox{e}^{-i k x_j}$) we obtain 
\begin{eqnarray}
 \hat{V}(\lambda) |x_j \rangle = \mbox{e}^{i \lambda x_j} |x_j \rangle \, , \,\,\,\,\, 
 \hat{U}(\delta) |x_j \rangle  = |x_j  - \delta \rangle.
\label{coordbase}
\end{eqnarray}
It is easy to verify that $\hat{V}(\lambda)$ is weakly continuous in $\lambda$, 
whence exists a self-adjoint operator $\hat{x}$ such that $\hat{x} |x_j \rangle = x_j |x_j\rangle$
\cite{AFW}, \cite{Fonte.Math}.

The operator analogy between loop quantum gravity and polymer representation is
the following: the basic operator of loop quantum gravity, holonomies and electric field fluxes, 
are respectively analogous to the operators $\hat{U}(\delta)$ and $\hat{x}$ with commutator 
$[\hat{x}, \hat{U}(\delta)] = - \delta \, \hat{U}(\delta)$.  The commutator is parallel to the commutator 
between electric fields and holonomies. 
As, in the polymer representation, the unitary operator $\hat{U}(\delta)$ is well defined but
the operator $\hat{p}$ doesn't exist, in loop quantum gravity the holonomies operators 
are unitary represented self-adjoint operators but the connection operator doesn't exist.
As $\hat{x}$, the electric flux operators are unbounded self-adjoint operators with discrete
eigenvalues. 

After this short review on polymer representation 
of the Weyl-Heisenberg algebra we return to our system.

\subsection{Polymer black hole quantization}

Following the previous section we 
quantize the Hamiltonian constraint and the inverse volume operator 
in the Polymer representation of the Weyl-Heisenberg algebra. 
The operators are $\hat{x}$, acting on the basis
states according to
\begin{eqnarray}
\hat{x} |\mu \rangle = L \mu |\mu \rangle \, , \,\,\,\,\,\,\, \langle \mu | \nu \rangle = \delta_{\mu \nu}
\label{xoperator}
\end{eqnarray}
(we have redefined the continuum eigenvalues of the position operator of the 
previous section $x_i \rightarrow \mu$)
and the operator corresponding to the classical
momentum function $U_{\delta} = e^{i \,8\pi G_N \delta p/L}$.  We define the
action of $\hat{U}_{\delta}$ on the basis states using the definition
(\ref{xoperator}) and using a quantum analog of the Poisson bracket
between $x$ and $U_{\delta}$
\begin{eqnarray}
\hat{U}_{\delta} |\mu \rangle = | \mu - \delta \rangle
\,\, ,   \,\,\,\,\,\,\, [ \hat{x} , \hat{U}_{\delta} ] = - \delta L \hat{U}_{\delta}.
\label{Traction}
\end{eqnarray}
Using the standard quantization procedure $[ \, , \, ] \rightarrow i
\hbar \{ \, , \, \}$, the Poisson bracket (\ref{Poisson.Volume}) and (\ref{Traction}) 
we obtain the value of the length scale $L = \sqrt{8 \pi \, G_N \hbar} = l_P$.

\subsubsection{Avoidance black hole singularity and regular dynamics}

We recall that the dynamics is all in the function $b(t)$, which is equal
to the the radial Schwarzschild coordinate inside the horizon. 
The important point is that $b(t = 0) = 0$ and this is the
Schwarzschild singularity.  We now show that the spectrum of 
the operator 
$\frac{1}{b(t)}$ does not diverge in quantum mechanics and therefore there 
is no singularity in the quantum theory.

Using the relation (\ref{unosux}),  
and promoting the Poisson brackets
to commutators, we obtain (for $\delta = 1$) the operator
\begin{eqnarray}
 \widehat{\frac{1}{|x|}} = \frac{1}{2 \pi G_N \hbar \, l_0^{\frac{2}{3}}} \left(
 \hat{U}^{-1} \left[ \hat{V}^{\frac{1}{3}} , \hat{U} \right] \right)^2. 
\end{eqnarray}
The action of this operator on the basis states is 
(the volume operator is diagonal on the basis states,  
$\hat{V} | \mu \rangle = l_0 |x|^{\frac{3}{2}} | \mu \rangle = l_0 |L
\mu|^{\frac{3}{2}} |\mu \rangle$) 
  \begin{eqnarray}
 \widehat{\frac{1}{|x|}} \, | \mu \rangle = \sqrt{\frac{2}{ \pi  G_N \hbar}}
 \left( | \mu |^{\frac{1}{2}} - |\mu -1|^{\frac{1}{2}}\right)^2 \, |
 \mu \rangle. 
\end{eqnarray}
We can now see that the spectrum is bounded from below and so we have
not singularity in the quantum theory.  In fact the
curvature invariant
$\mathcal{R}_{\mu \nu \rho \sigma} \, \mathcal{R}^{\mu \nu \rho \sigma}
= 48 M^2 G_N^2/x(t)^6$
is finite in quantum mechanics in $\mu = 0$. The eigenvalue of the operator $1/|x|$ for
the state $|0 \rangle$ corresponds to the classical singularity and in
the quantum case it is $4/l_P^2$, which is the largest
possible eigenvalue.  For this particular value the curvature
invariant it is not infinity
 \begin{eqnarray}
\widehat{\mathcal{R}_{\mu \nu \rho \sigma} \, \mathcal{R}^{\mu \nu
\rho \sigma}} |0 \rangle = \widehat{\frac{48 M^2 G_N^2}{|x|^6}} |0
\rangle = \frac{384 M^2 G_N^2}{\pi^3 l_P^6} |0 \rangle. 
\end{eqnarray}
If we consider the $\hbar \rightarrow 0$ limit we obtain the classical 
singularity so the result is a genuinely quantum gravity effect). 
On the other hand, for $|\mu | \rightarrow \infty$ the eigenvalues go
to zero, which is the expected behavior of $1/|x|$ for large $|x|$.

Now we study the quantization of the Hamiltonian constraint near the
singularity, in the approximation (\ref{Hamiltonian.approx}).  There
is no operator $p$ in polymer quantum representation that we have chosen,
hence we choose the following alternative representation for $p^2$. 
Consider the classical expression
\begin{eqnarray}
p^2 = \frac{L^2}{(8 \pi G_N)^2} \lim_{\delta \rightarrow 0}
\left( \frac{2 - U_{\delta} - U_{\delta}^{-1}}{\delta^2} \right).
\label{p}
\end{eqnarray}
We have can give a physical interpretation to $\delta$ as $\delta =
l_p / L_{phys}$, where $L_{Phys}$ is the characteristic size of the
system. Using (\ref{p}) we write the Hamiltonian constraint as
\begin{eqnarray}
\hat{H} = \frac{\mathcal{C}}{l_0^{1/3} L^{1/2}} \left[ \hat{U}_{\delta} +
\hat{U}_{\delta}^{-1} - (2 - \mathcal{C}^{\prime}) \, \hat{\mathbb{I}} \, \right] \mbox{sgn}(x) \left(
\hat{U}^{-1} \left[ \hat{V}^{\frac{1}{3}} , \hat{U} \right] \right)
\end{eqnarray}
where $\mathcal{C} = \frac{L^{5/2} \, G_N}{{(8 \pi G_N})^{5/2} \delta ^3 \, R \, l_0^{1/3} \hbar }$
and $\mathcal{C}^{\prime} = \frac{8 \pi \, R^2 \delta^2}{ l_P^2}$.  The action of $\hat{H}$
on the basis states is
\begin{eqnarray}
&& \hat{H} | \mu \rangle = \mathcal{C} \, \mathcal{V}(\mu)
\left[ | \mu - \delta \rangle + | \mu + \delta \rangle - (2 -
\mathcal{C}^{\prime}) | \mu \rangle \right], \nonumber \\
&& \hspace{-0.1cm}\mathcal{V}(\mu) = \left\{ \begin{array}{cc} - \left|
|\mu - \delta |^{1/2} - |\mu|^{1/2} \right| \hspace{0.5cm} \mbox{for}
\hspace{0.5cm} \mu \neq 0 \\
 |\delta|^{1/2} \hspace{3cm} \mbox{for} \hspace{0.5cm} \mu = 0 \\
 \end{array}\right.
\label{sistem3}
\end{eqnarray}

We now calculate the solutions of the the Hamiltonian constraint.  The
solutions are 
in the dual space of 
$H_{Poly}$.  A generic element of this space is
$\langle \psi | = \sum_{\mu} \psi(\mu) \langle \mu |$. 
The constraint equation $\hat{H} |\psi \rangle = 0$ is now interpreted
as an equation in the dual space $\langle \psi | \hat{H}^{\dag}$; from
this equation we can derive a relation for the coefficients
$\psi(\mu)$
\begin{eqnarray}
\mathcal{V}(\mu + \delta) \, \psi(\mu + \delta) +
\mathcal{V}(\mu - \delta) \, \psi(\mu - \delta) - (2 -
\mathcal{C}^{\prime})\, \mathcal{V}(\mu) \, \psi(\mu) =
0.
\label{difference1}
\end{eqnarray}
This relation determines the coefficients for the physical dual state. 
We can interpret this states as describing the $quantum$ $spacetime$
near the singularity.  From the difference equation (\ref{difference1})
we obtain physical states as combinations of a countable number of
components of the form $\psi(\mu + n \delta) |\mu + n \delta \rangle $
($\delta \sim l_P/L_{Phys} \sim 1$); any component corresponds to a
particular value of volume, so we can interpret $\psi(\mu + \delta)$
as the wave function describing the black hole near the singularity at
the time $\mu + \delta$.  A solution of the Hamiltonian constraint
corresponds to a linear combination of black hole states for
particular values of the volume or equivalently particular values of
the time.

\section{Loop quantum black hole}
In this section we quantize the Kantowski-Sachs space-time in Ashtekar 
variables and without approximations \cite{ABM}. 

\subsection{Ashtekar variables for Kantowski-Sachs space-time} 
The Kantowski-Sachs space-time is a simplified version of an 
homogeneous but anisotropic spacetime, written in coordinates $(t,r,\theta, \phi)$. 
An homogeneous but anisotropic space-time of spatial section $\Sigma$
of topology $\Sigma \cong \mathbb{R} \times S^2$ is characterized by an invariant 
connection 1-form $A_{[1]}$ of the form \cite{BojThiemann}, \cite{BojImp}
\begin{eqnarray}
A_{[1]} = A_r(t) \, \tau_3 \, dr + (A_1(t) \, \tau_1 + A_2(t) \tau_2) \, d \theta +
           (A_1(t) \, \tau_2 - A_2(t) \tau_1) \sin \theta \, d \phi + \tau_3 \, \cos \theta \, d \phi.
           \label{symconnection}
\end{eqnarray} 
The $\tau_i$ are the generators of the $SU(2)$ fundamental representation. 
They are related to the
Pauli $\sigma_i$ matrix by $\tau_i = -  \frac{i}{2} \sigma_i$.
On the other side the dual invariant densitized triad is 
\begin{eqnarray}
E_{[1]} = E^r(t) \, \tau_3 \, \sin \theta \, \frac{\partial}{\partial r} + (E^1(t) \, \tau_1 + E^2(t) \, \tau_2)
\, \sin \theta \, \frac{\partial}{\partial \theta}  +
           (E^1(t) \, \tau_2 - E^2(t) \tau_1) \frac{\partial}{\partial \phi}.
           \label{symtriad}
\end{eqnarray} 
Since spacetime is homogeneous, the diffeomorphism constraint is
automatically satisfied. 

In this paper we study the Kantowski-Sachs space-time 
with space section of topology $\mathbb{R} \times S^2$; the connection $A_{[1]}$ 
is more simple than in (\ref{symconnection}) with $A_2 = A_1$, and in the triad 
(\ref{symtriad}) we can choose the gauge $E^2 = E^1$ \cite{Bombelli}. There is 
a residual gauge freedom on the pair $(A_1, E^1)$. This is a discrete transformation
$P: (A_1, E^1) \rightarrow (-A_1, -E^1)$; we have to fix this symmetry on the Hilbert space.
The Gauss constraint 
is automatically satisfied and the Euclidean part of the Hamiltonian constraint becomes
\begin{eqnarray}
\mathcal{H}_E = \frac{2 \sqrt{2} \sin \theta \,\,\mbox{sgn}(E^r)}{\sqrt{|E^r|}|E^1|} \, 
\Big[2 A_r E^r A_1 E^1 + ( 2 (A_1)^2  - 1)(E^1)^2 \Big]
 \label{hamiltonian}
\end{eqnarray}
we redefine $E^r \equiv E$ and $A_r \equiv A$. 
The connection between the metric (\ref{KS}) and the density triad is 
$q_{a b} = \mbox{diag}(
                          2 (E^1)^2/|E|,
                            |E|,  |E| \, \sin^2 \theta)$.
 Another useful quantity is  
the volume of the spatial section $\Sigma$
\begin{eqnarray}
V = \int dr \, d \phi \, d \theta \, \sqrt{\mbox{det}(q)} = 4 \pi \sqrt{2} R \sqrt{|E|} |E^1|,
\label{Volume}
\end{eqnarray}
where $R$ is a cut-off on the space radial coordinate. 
The spatial homogeneity enable us to fix a linear radial cell $\mathcal{L}_r$ and restrict all 
integrations to this cell \cite{Boj}.
To simplify notations we restrict the linear radial cell to the Planck length $l_P$  and so we can take $\int dr = R \equiv l_P$ in the action functional (\ref{Volume}).
Now we are going to use $R \equiv l_P$ in all the paper.

The classical symplectic structure of the phase space can be obtained by inserting the symmetry reduced variables in the action (\ref{action}). This gives
\begin{eqnarray}
 S =  \frac{1}{\kappa \, \gamma} \int d t \int d r \, d \phi \, d \theta \, \sin \theta \,
\left[\mbox{Tr} \left( -2 E^a \dot{A}_{a} \right) + \dots \right] 
=  \frac{4 \pi l_P}{\kappa \, \gamma} \int d t  \, \left[E \dot{A} + 4 E^1 \dot{A}_1 + \dots \right].
\label{actionred}
\end{eqnarray}

We can read the symplectic structure
of the classical phase space directly from the reduced action (\ref{actionred}). The phase space consists of two canonical  pairs $A, E$ and $A_1, E^1$ and from (\ref{actionred}) we can obtain
the simplectic structure. The simplectic structure is given by 
the poisson brackets, $\{A, E\} = \frac{\kappa \gamma}{4 \pi l_P}$ and 
$\{A_1, E^1\} = \frac{\kappa \gamma}{16 \pi  l_P}$.  
The coordinates and the momenta have dimensions: $[A] = L^{-1}$, $[A_1] = L^0$,
$[E] = L^2$ and $[E^1] = L$. 

The elementary configuration variables used in LQG are given by the holonomies along curves in 
the spatial section $\mathbb{R} \times S^2$ and the fluxes of triads on a two-surface in
$\mathbb{R} \times S^2$.
We restrict our attention to three sets of curves. 
More precisely, we consider  only spin networks \cite{book} 
based on graph made just of radial edges, 
and of edges along circles in the $\theta$-direction or at $\theta = \pi/2$.

Let us introduce the fiducial triad $^o e^a_I = \mbox{diag}(1, 1, \sin^{-1} \theta)$ and co-triad 
$^o \omega^I_a = \mbox{diag}(1, 1, \sin \theta)$. The holonomy along a curve
in the direction ``$I \, $" is given by $h_I=\exp\int A^i_I \tau_i$,
\begin{eqnarray}
h_1 = 
\exp[ A \mu_0 l_P \, \tau_3] \,, \,\,\,\, 
 h_2 = 
 \exp [A_1 \mu_0  \, (\tau_2 + \tau_1)] \, , \,\,\,\, 
 h_3 = 
 \exp [A_1 \mu_0   \, (\tau_2 - \tau_1) ] \, , 
\label{holonomiyI}
\end{eqnarray}
where $A_1^i = (0, 0, A)$, $A_2^i = (A_1, A_1,0)$ and $A_3^i = (-A_1, A_1, 0)$.
The connection in (\ref{holonomiyI}) is integrated in the direction ``$I \, $";
$\mu_0 l_P$ is the length of the curve along the direction $r$, 
$\mu_0$ is the length of the curve along the directions $\theta$ and  $\phi$. 
The length are defined using the fiducial triad  $^o e^a_I$.

Recall that the Hamiltonian constraint can be written in terms of the curvature
$F_{ab}$ and the Poisson bracket between $A_a$ and the volume $V$ \cite{Thie}. 
The Euclidean part of the Hamiltonian constraint becomes   
\begin{eqnarray}
H_E = - \frac{4}{\kappa \gamma} \int d^3 x \, N \, \epsilon^{a b c} \, 
\mbox{Tr} \left[F_{a b} \, \{A_{c}, V\} \right].
\label{hamiltonianE}
\end{eqnarray}
Because of homogeneity we can assume that the lapse function $N$ is constant and 
in the rest  of the paper we will set $N=1$.

We can express the curvature $F_{ab}$ and the Poisson bracket $\tau_i \, \{ A^i_a, V \}$ 
in terms of holonomies \cite{ABM} 
obtaining the following form for the Hamiltonian constraint 
\begin{eqnarray}
 H_E  =  - \frac{16 \pi}{\kappa \gamma \mu_0^3} \, \sum_{I J K}\, \epsilon^{I J K} \, \mbox{Tr} \left[h_I h_J h_I^{-1} h_J^{-1} h_{[IJ]} \, h_K^{-1} \{h_K, V \}  \right],
\label{hamiltonianEreg2}
\end{eqnarray}
where $h_{[IJ]} = \exp(- \mu_0^2 \, C_{I J} \, \tau_3)$ and $C_{IJ} = \delta_{2 I} \delta_{3 J} - \delta_{3_I} \delta_{2 J}$.

Using the classical identity 
 \begin{eqnarray}
 \frac{\mbox{sgn}(E)}{(\mbox{det}(q_{ab}))^{\frac{1}{4}}} = - \frac{64 \, (4 \pi l_P)^{\frac{3}{2}}}{3! \, \kappa^3 \gamma^3(\sin \theta)^{\frac{3}{2}}} \,
 \epsilon_{i j k} \, \epsilon^{a b c} \, 
 \{A^i_a, V^{\frac{1}{2}} \}
  \{A^j_b, V^{\frac{1}{2}} \}
   \{A^k_c, V^{\frac{1}{2}} \},
   \label{ClassicVol}
      \end{eqnarray} 
we can define the inverse of the volume in terms of holonomies. The result is \cite{ABM} 
\begin{eqnarray}
&&\hspace{-0.7cm} \frac{\mbox{sgn}(E)}{(\mbox{det}(q))^{\frac{1}{4}}}  =  
\frac{256  (4 \pi)^{\frac{3}{2}} \sqrt{l_P}}{3  \kappa^3  \gamma^3 \mu_0^3 (\sin \theta)^{\frac{1}{2}}}  \, 
 \epsilon_{i j k} \, \sum_{I J K} \epsilon^{I J K} \, 
 \mbox{Tr}[\tau^i h_I^{-1} \{h_I, V^{\frac{1}{2}} \}] \, 
  \mbox{Tr}[\tau^j h_J^{-1} \{h_J, V^{\frac{1}{2}} \}] \,
  \mbox{Tr}[\tau^k h_K^{-1} \{h_K, V^{\frac{1}{2}} \} ]. \nonumber \\
 &&
    \label{ClassicVol2}
      \end{eqnarray}
The definition (\ref{ClassicVol2}) will be useful to calculate the inverse 
volume spectrum.
\subsection{Quantum Theory}
We construct the kinematical Hilbert space $H_{kin}$ for the 
Kantowski-Sachs minisuperspace model in analogy with the full theory. 
As in the second section we define a graph $\Gamma$ as a finite number of couple of points 
$(\mu_{E i} \, , \, \mu_{E^{1} i})$, where $\mu_{E i}, \mu_{E^1i} \in \mathbb{R}$.
We denote by $\mbox{Cyl}_{\Gamma}$ the vector space of function $f(A, A_1)$ 
($f : \mathbb{R}^2 \rightarrow  \mathbb{C}$) of the type 
\begin{eqnarray}
f(A, A_1) = \sum_{i \, j} f_{ij} \,
e^{\frac{i \mu_{E i} \, l_P \, A}{2} + \frac{i \mu_{E^1 j} \, A_1}{\sqrt{2}}}.
\label{genfun}
\end{eqnarray}
where $A, A_1 \in \mathbb{R}$, $\mu_{E i} \, , \, \mu_{E^1 j} \in \mathbb{R}$ ,
$f_{ij} \in \mathbb{C}$ and 
$i,j$ run over a finite number of integers (labelling the points of the graph). We call
the function $f(A, A_1)$ in Cyl$_{\Gamma}$ cylindrical  with respect to the graph $\Gamma$.
We consider all possible graphs (the points and their number can vary 
from a graph to another) and denote by Cyl the infinite dimensional 
vector space of functions cylindrical with respect to some graph: $\mbox{Cyl}=\bigcup_{\Gamma}\mbox{Cyl}_{\Gamma}$. 
Thus, any element $f(A, A_1)$ of Cyl can be expanded as in (\ref{genfun}),
where the uncountable basis 
$e^{\frac{i \mu_{E i} \, l_P \, A}{2}} \otimes e^{\frac{i \mu_{E^1 j} \, A_1}{\sqrt{2}}}$
is now labeled by arbitrary real numbers $(\mu_{E} \, , \, \mu_{E^{1}})$.
A basis in Cyl is given by $|\mu_E,  \mu_{E^1} \rangle \equiv |\mu_E \rangle \otimes |\mu_{E^1} \rangle$.  Introducing the standard bra-ket notation we can define a basis \cite{MAT} in the
Hilbert space via  
\begin{eqnarray}
\langle A|\mu_E\rangle \otimes \langle A_1|\mu_{E^1} \rangle =  e^{\frac{i \mu_E \, l_P \, A}{2}} \otimes e^{\frac{i \mu_{E^1} \, A_1}{\sqrt{2}}}.
\label{basis}
\end{eqnarray} 
The basis states (\ref{basis}) are normalizable in contrast to the standard quantum mechanical 
representation and they satisfy 
\begin{eqnarray}
\langle \mu_E, \mu_{E^1}| \nu_E, \nu_{E^1} \rangle = \delta_{\mu_E, \nu_E} \, \delta_{\mu_{E^1}, \nu_{E^1}}.
\end{eqnarray}
The Hilbert space $\mathcal{H}_{kin}$ is the Cauchy completion of Cyl or more succinctly $\mathcal{H}_{kin} = L_2(\bar{\mathbb{R}}^2_{Bohr}, d \mu_0)$, where $\bar{\mathbb{R}}_{Bohr}$ is the Bohr-compactification of $\mathbb{R}$ and $d\mu_0$ is the Haar measure on $\bar{\mathbb{R}}^2_{Bohr}$. 
In LQG (or in polymer representation)
the fundamental operators are $\hat{E}, \hat{E}^1, \hat{h}_I$ and 
\begin{itemize}
\item the momentum operators can be represented on the Hilbert space by 
\begin{eqnarray}
\hat{E} \rightarrow - i \frac{\gamma \, l_P}{4 \pi} \, \frac{d}{d A} \, ,  \,\,\,\,\,\,\,\,
\hat{E}^1 \rightarrow - i \frac{\gamma \, l_P}{16 \pi} \, \frac{d}{d A_1},
\label{EE1}
\end{eqnarray}
and the he spectrum of these two momentum operators on the Hilbert space 
basis is 
\begin{eqnarray}
\hat{E} |\mu_E, \mu_{E^1} \rangle = \frac{\mu_E \,  \gamma \, l_P^2}{8 \pi} |\mu_{E}, \mu_{E^1} \rangle \, ,
\,\,\,\,\,\,\,\, \hat{E}^1 |\mu_E, \mu_{E^1} \rangle = \frac{\mu_{E^1} \, \gamma \, l_P}{16 \pi \sqrt{2}} |\mu_E, \mu_{E^1} \rangle;
\label{diagEE1}
\end{eqnarray}
\item  the holonomy operators $\hat{h}_{I}$ in the directions $r$, $\theta$, $\phi$ of the space 
 section $\mathbb{R} \times S^2$ are : $\hat{h}_1^{(\mu_E)}$, $\hat{h}_2^{(\mu_{E^1})}$ and 
 $\hat{h}_3^{(\mu_{E^1})}$,
 where $\mu_{E} l_P$ is the length along the radial direction $r$ 
and $\mu_{E_1}$ is the length along the directions $\theta$ and $\phi$
(all the length are define using the fiducial triad $^o e^a_I$). The holonomies 
operators act on the Hilbert space $\mathcal{H}_{kin}$ by multiplication.

\end{itemize}

 We have to fix the residual gauge freedom on the Hilbert space. We consider the
 operator $\hat{P} : |\mu_E, \mu_{E^1} \rangle \rightarrow |\mu_E, - \mu_{E^1} \rangle$
 and we impose that only the invariant states (under $\hat{P}$) are in the kinematical
 Hilbert space. The states in the Hilbert space are : 
 $\frac{1}{\sqrt{2}} \left[|\mu_E, \mu_{E^1} \rangle + |\mu_E, - \mu_{E^1} \rangle\right]$
 for $\mu_{E^1} \neq 0$ and the states $| \mu_E , 0 \rangle$ for $\mu_{E^1} = 0$.

\subsubsection{Inverse volume spectrum}
In this section we  study the black 
hole singularity problem in loop quantum gravity 
calculating the spectrum of the inverse volume operator.
 The operator version of the quantity $\mbox{sgn}(E)/(\mbox{det} (q))^{\frac{1}{4}}$
 defined in the formula (\ref{ClassicVol2}) is 
 \begin{eqnarray}
&& \hspace{-0.6cm}\widehat{\frac{\mbox{sgn}(E)}{(\mbox{det}(q))^{\frac{1}{4}}}} =  
\frac{256 \, i (4 \pi l_P)^{\frac{3}{2}} \epsilon_{i j k} }{3 \, l_P^7 \gamma^3 \mu_0^3 (\sin \theta)^{\frac{1}{2}}}  
\sum_{I J K} \epsilon^{I J K} \, \mbox{Tr}\left[\tau^i \hat{h}_I^{-1} [\hat{h}_I, \hat{V}^{\frac{1}{2}}] \right] \,  \mbox{Tr}\left[\tau^j \hat{h}_J^{-1} [\hat{h}_J, \hat{V}^{\frac{1}{2}}] \right] \,\mbox{Tr}\left[\tau^k \hat{h}_K^{-1} [\hat{h}_K, \hat{V}^{\frac{1}{2}}]
\right].    \nonumber \\
&&
\label{ClassicVol2Q}
      \end{eqnarray} 
To calculate the action of  (\ref{ClassicVol2Q}) on the Hilbert space basis
we Introduce the normalized vectors $n_1^i = (0, 0, 1)$, $n_2^i = \frac{1}{\sqrt{2}} (1, 1, 0)$,
  $n_3^i = \frac{1}{\sqrt{2}} (-1, 1, 0)$. 
  Using the vectors $n^i_I$ 
  we can rewrite the holonomies $h_I$ 
  of (\ref{holonomiyI}) as
$ h_I= \exp ( \bar{A}_I \, n^i_I  \, \tau_i )= \cos (\bar{A}_I/2 ) + 2 n^i_I \, \tau_i \, \sin (\bar{A_I}/2 )$, 
 where $\bar{A}_{I=1} = A l_P \mu_0$ and $\bar{A}_{I=2} = \bar{A}_{I=3} = A_1 \mu_0 \sqrt{2}$.
 The action of the multiplicative operator $\hat{h}_I$ on the bases states is 
 \begin{eqnarray}
 \hat{h}_I |\mu_I \rangle = \frac{\mathbb{I} -i n^i_I \tau_i}{2} |\mu_I + \mu_0 \rangle + 
        \frac{\mathbb{I}+ i n^i_I \tau_i}{2} |\mu_I - \mu_0 \rangle.
        \label{holAct}
 \end{eqnarray}
The spectrum of the inverse volume operator is 
\begin{eqnarray}
&& \widehat{\frac{1}{(\mbox{det}(q))^{\frac{1}{4}}}} 
|\mu_E, \mu_{E^1} \rangle =
\Big(\frac{2 \pi}{\gamma}\Big)^{\frac{3}{4}} \frac{8}{l_P \, \mu_0^3 \, (\sin \theta)^{\frac{1}{2}}} \, 
|\mu_E|^{\frac{1}{2}} \, |\mu_{E^1}|^{\frac{1}{2}} \nonumber \\
&&\hspace{2.5cm}
 \left| |\mu_E + \mu_0|^{\frac{1}{4}} - 
|\mu_E - \mu_0|^{\frac{1}{4}}\right| \, \left(|\mu_{E^1} + \mu_0|^{\frac{1}{2}} - 
|\mu_{E^1} - \mu_0|^{\frac{1}{2}}\right)^2|\mu_E, \mu_{E^1} \rangle.
   \label{spectrum} 
     \end{eqnarray}

\begin{figure}
 \begin{center}
  \includegraphics[height=5.5cm]{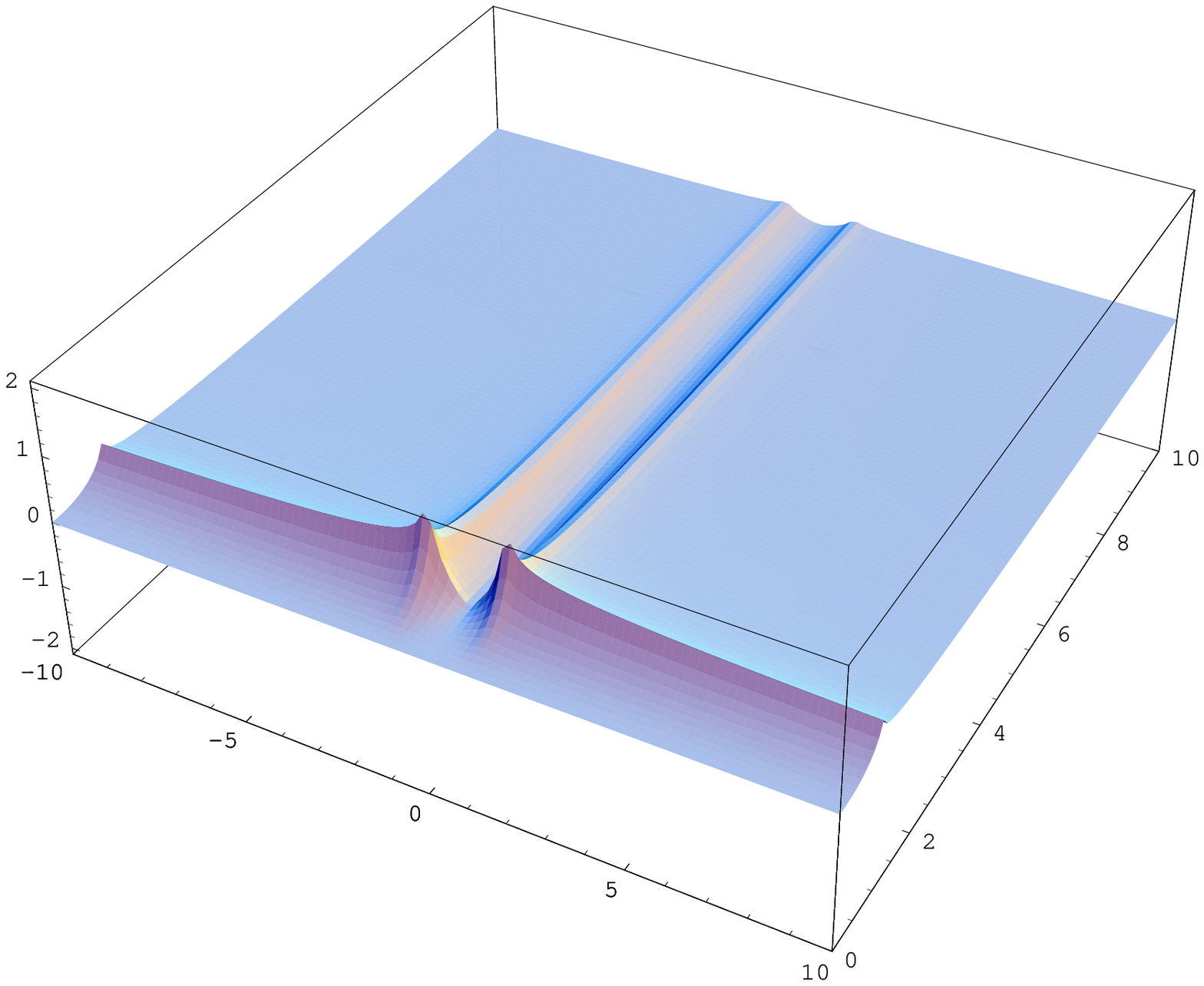} \,\,
   \includegraphics[height=5.5cm]{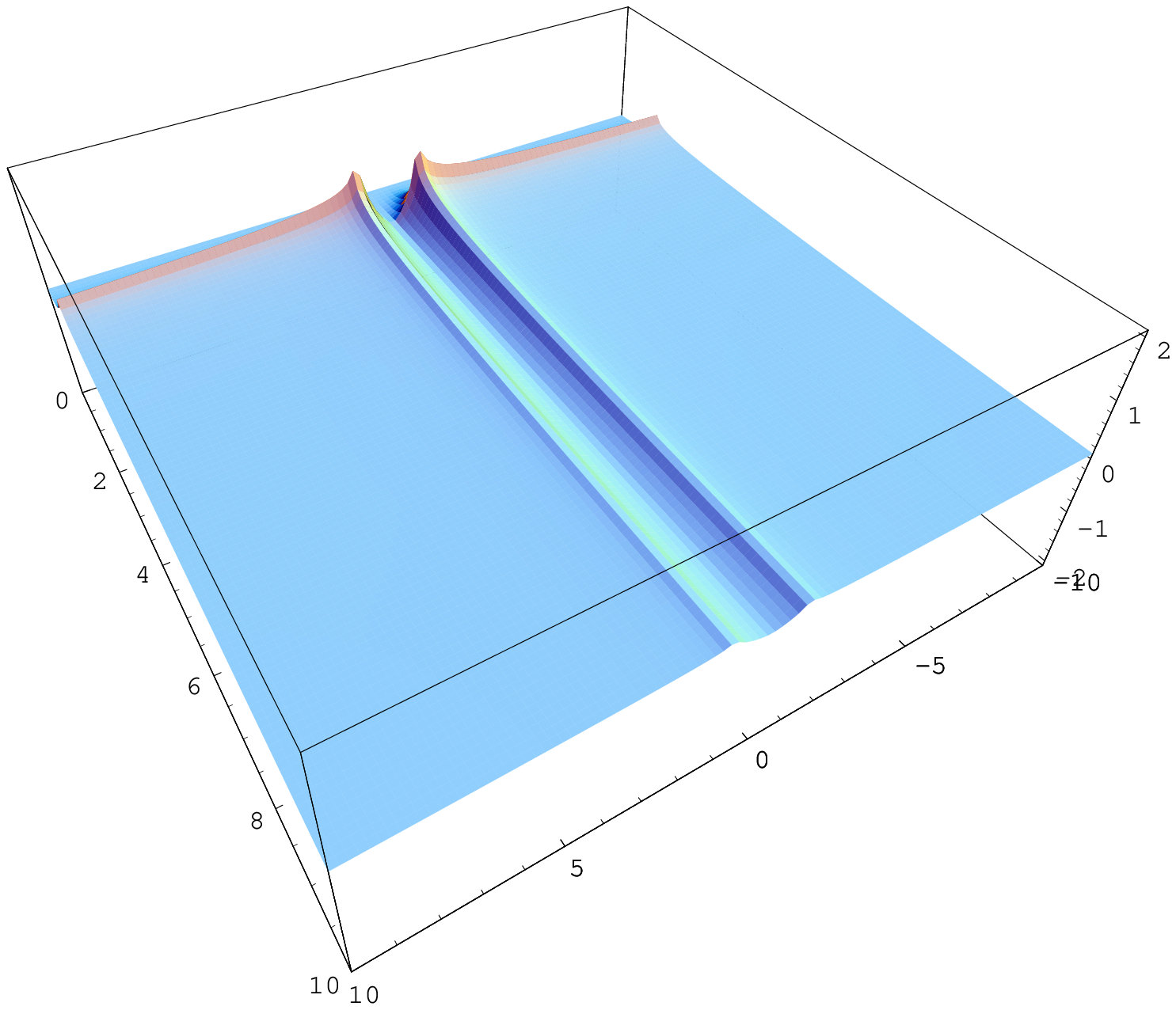}
  \end{center}
  \caption{\label{InVol} 
                 Plot of the inverse volume spectrum for $\mu_{E^1} \geqslant 0$ and $\forall \mu_E$.}
  \end{figure}

The spectrum of the inverse volume operator is bounded above, near the classical
singularity which is in $E=0$ or $\mu_E=0$, and reproduces the correct classical 
spectrum of $1/(\mbox{det}(q))^{\frac{1}{4}}$ for large volume eigenvalues (Fig.\ref{InVol}).

\subsubsection{Quantum dynamics} 

In this section we study the dynamics of the model solving the Hamiltonian
constraint in (the dual space of) the Hilbert space.

The quantum version of the Hamiltonian constraint 
defined in (\ref{hamiltonianEreg2}) can be obtained promoting the classical holonomies
to operators and the poisson bracket to the commutator    
\begin{eqnarray}
 \hat{H}_E  =  \frac{16 \pi \, i}{\mu_0^3 \, \gamma \, l_P^2} \, \sum_{I J K}\, \epsilon^{I J K} \, \mbox{Tr} \left[ \hat{h}_I \hat{h}_J \hat{h}_I^{-1} \hat{h}_J^{-1} \hat{h}_{[IJ]} \, \hat{h}_K^{-1} [\hat{h}_K, \hat{V} ]  \right].
\label{hamiltonianEreg2Q}
\end{eqnarray}
Using the relations in (\ref{holAct})
we can calculate the action of the Hamiltonian constraint 
on the Hilbert space basis  $| \mu_E, \mu_{E^1} \rangle$ and
solve the Hamiltonian constraint
to obtain the physical states.
As in non-trivially constrained systems, we expect that the physical states are not 
normalizable in the kinematical Hilbert space. However, as in the full loop quantum 
gravity, we again have the Gelfand triple  
$\mbox{Cyl} \subset \mathcal{H}_{kin} \subset \mbox{Cyl}^*$
and the physical states will be in $\mbox{Cyl}^*$, which is the algebraic dual of 
$\mbox{Cyl}$. 
An element of this space is 
$\langle \psi | = \sum_{\mu_E, \mu_{E^1}} \psi_{\mu_E, \mu_{E^1}} \langle \mu_E, \mu_{E^1}|$.
The constraint equation $\hat{H} |\psi \rangle = 0$ is now interpreted as an equation in the dual space $\langle \psi | \hat{H}^{\dag}$;
from this equation we can derive a relation for the coefficients $\psi_{\mu_E, \mu_{E^1}}$  
\begin{eqnarray}
&& \hspace{-0.7cm}- \alpha_{\mu_E - 2 \mu_0, \mu_{E^1} - 2 \mu_0} \, \psi_{\mu_E - 2 \mu_0, \mu_{E^1} - 2 \mu_0} 
+ \alpha_{\mu_E + 2 \mu_0, \mu_{E^1} - 2 \mu_0} \, \psi_{\mu_E + 2 \mu_0, \mu_{E^1} - 2 \mu_0} \nonumber \\
&&\hspace{-0.7cm} + \alpha_{\mu_E - 2 \mu_0, \mu_{E^1} + 2 \mu_0} \, \psi_{\mu_E - 2 \mu_0, \mu_{E^1} + 2 \mu_0} 
- \alpha_{\mu_E + 2 \mu_0, \mu_{E^1} + 2 \mu_0} \, \psi_{\mu_E + 2 \mu_0, \mu_{E^1} + 2 \mu_0} \nonumber \\
&& \hspace{-0.7cm}+ \frac{\sin(\mu_0^2/2) -  \cos(\mu_0^2/2)}{2}
\left(\beta_{\mu_E, \mu_{E^1} - 4 \mu_0} \, \psi_{\mu_E, \mu_{E^1} - 4 \mu_0} -
2 \beta_{\mu_E, \mu_{E^1}} \, \psi_{\mu_E, \mu_{E^1}} 
 + \beta_{\mu_E, \mu_{E^1} + 4 \mu_0} \, \psi_{\mu_E, \mu_{E^1} + 4 \mu_0}\right) \nonumber \\
&& \hspace{-0.7cm} - 2 \sin(\mu_0^2/2) \left(\beta_{\mu_E, \mu_{E^1} - 2 \mu_0} \, \psi_{\mu_E, \mu_{E^1} - 2 \mu_0} +
\beta_{\mu_E, \mu_{E^1} +2 \mu_0} \, \psi_{\mu_E, \mu_{E^1} +2 \mu_0}\right) = 0,
\label{solution}
\end{eqnarray}
where the functions $\alpha(\mu_E, \mu_{E^1})$ and  $\beta(\mu_E, \mu_{E^1})$
are define by 
\begin{eqnarray}
\alpha_{\mu_E, \mu_{E^1}} := |\mu_E|^{\frac{1}{2}}
\left(|\mu_{E^1} + \mu_0| - |\mu_{E^1} -\mu_0| \right) \,\, , \,\,\,\,\,
\beta_{\mu_E, \mu_{E^1}} := |\mu_{E^1}|
(|\mu_{E} + \mu_0|^{\frac{1}{2}} - |\mu_{E} -\mu_0|^{\frac{1}{2}} )
\label{alfabeta}
\end{eqnarray}
How can be seen from equation (\ref{solution}) the quantization program 
produce a difference equation and imposing a boundary condition
we can obtain the wave function $\psi_{\mu_E, \mu_{E^1}}$ for the black hole.
We can interpret $\psi_{\mu_E, \mu_{E^1}}$ as the wave function of the anisotropy 
 ``$E^1$" at the time  ``$E$". 
  It is evident from (\ref{alfabeta}) that the dynamics is regular in $\mu_E=0$ where the
  classical singularity is localized. 
 As in loop quantum cosmology also in this case the state $\psi_{0,0}$ decouples from the dynamics and the quantum evolution does not stop at the classical singularity.
The ``other side" of the singularity corresponds to a new domain where
the triad reverses its orientation.

\section*{Conclusions}
In this paper we have
summarized loop quantum gravity theory and we have applied the 
ideas to study the space-time region inside the Schwarzschild black hole 
horizon. Because the space-time region inside the horizon is spatially homogeneous of 
Kantowski-Sachs type \cite{KS}, 
we have studied this minisuperspace model. 
This is an homogeneous but anisotropic minisuperspace model
with spatial topology $\mathbb{R} \times S^2$. 
We have analyzed the model firstly in ADM variables with some drastic 
simplification and then 
in Ashtekar variables. 
We have quantized the reduced model 
using a quantization procedure induced by the full ``loop quantum gravity".
Our analysis it was useful 
in order to understand what happens close to the black hole singularity where 
quantum gravity effects are dominant and the classical Schwarzschild solution 
is not correct. 
 
The main results are :

\begin{enumerate}
\item  the curvature invariant and 
 the inverse volume operator have 
a finite spectrum in all the region
inside the horizon and we can conclude that the classical singularity disappears 
at the kinematical level;
on the other side
for large eigenvalues of the volume operator 
we find the classical inverse volume behavior,
\item  the solution of the Hamiltonian constraint gives a difference
equation for the coefficients of the physical states defined in the 
dual space of some dense subspace of the kinematical Hilbert space.
All the coefficients in the difference equation are regular in the classical
singular point then we have a solution of the singularity problem
also at the dynamical level. 
\end{enumerate}

An important consequence of the quantization is that, unlike the classical 
evolution, the quantum evolution does not stop at the classical singularity
and the ``other side" of the singularity corresponds to a new domain where
the triad reverses its orientation. 
From the difference equation 
we obtain 
physical states as combinations of a countable number of  components of the form
$\psi_{\mu_E + n \, \mu_0, \mu_{E^1} + m \, \mu_0} |\mu_E + n \, \mu_0, \mu_{E^1} + m \, \mu_0 \rangle$ (where $\mu_0 
\sim 1$ at the Plank scale and $n, m \in \mathbb{Z}$); any component corresponds to a particular value of the volume of the space section. We can interpret $\mu_E$ as the time and the anisotropy $\mu_{E^1}$ as the space partial observable \cite{Partial} that defines the quantum fluctuations around the Schwarzschild solution. We recall that $|E| = b^2$, therefore 
in the classical theory and in quantum mechanics, we can 
regard $|E|$ as an internal time. 
So the function $\psi_{\mu_E + n \, \mu_0, \mu_{E^1} + m \, \mu_0}$
is the wave function of the Black Hole inside the horizon 
at the time $\mu_E + n \, \mu_0$ and we have a natural and regular 
evolution beyond the classical singularity point which is in $\mu_E=0$ localized.  
A solution of the Hamiltonian constraint corresponds to a linear combination of 
black hole states for particular values of the anisotropy $\mu_{E^1}$ at the time $\mu_E$.

{\em It is interesting to recall that beyond the classical 
singularity the eigenvalue $\mu_E$ is negative and so we can suggest 
a new universe was born 
from the black hole formation process. 
In LQBH scenario pure states which fall into black hole emerge in a new universe as pure states
and the information loss problem is avoided. Information is not lost in the black hole
but it exists again in the space-time region in the future of the avoided singularity}.

\section*{Acknowledgements}
I am grateful to Carlo Rovelli, Eugenio Bianchi and Roberto Balbinot.


\begin{thebibliography}{99}


\bibitem{book}
Carlo Rovelli, {\em Quantum Gravity}, (Cambridge University Press,
Cambridge, 2004);
Alejandro Perez, {\em Introduction to loop quantum gravity and spin foams}
gr-qc/0409061;
T. Thiemann, {\em Loop quantum gravity: an inside view}, hep-th/0608210; 
T. Thiemann, {\em Introduction to Modern Canonical Quantum General Relativity},
gr -qc/0110034; {\em Lectures on Loop Quantum Gravity},
Lect. Notes Phys. 631, 41-135 (2003), gr-qc/0210094;
Alejandro Perez, {\em Spin foam models for quantum gravity}, 
Class. Quant. Grav. 20, R43 (2003), gr-qc/0301113 
  

\bibitem{MR} Martin Reuter, {\em Non perturbative
evolution equation for quantum gravity}, Phys. Rev. D57 971-985 (1998), hep-th/9605030

\bibitem{MV} F. Conrady, L. Doplicher, R. Oeckl, C. Rovelli and M. Testa
{\em Minkowski vacuum in background independent quantum gravity}, Phys. Rev. D 69 064019, 
gr-qc/0307118; 
Daniele Colosi, Luisa Doplicher, Winston Fairbairn, Leonardo Modesto,
Karim Noui and Carlo Rovelli, {\em Background independence in a nutshell: dynamics of a tetrahedron}, Class. Quant. Grav. 22 (2005) 2971-2989, gr-qc/0408079 

\bibitem{ModestoRovelli} Leonardo Modesto and Carlo Rovelli, {\em Particle scattering
in loop quantum gravity}, Phys. Rev. Lett. 95 191301 (2005), gr-qc/0502036 

\bibitem{Rovelli1} Carlo Rovelli, {\em Graviton propagator from background-independent 
quantum gravity}, Phys. Rev. Lett. 97 151301 (2006), gr-qc/0508124

 \bibitem{BMSR} Eugenio Bianchi, Leonardo Modesto, Carlo Rovelli and Simone Speziale,
 {\em Graviton propagator in loop quantum gravity}, Class. Quant. Grav. 23 (2006) 6989 -7028,
 gr-qc/0604044
 
 \bibitem{Simone12} Simone Speziale, {\em Towards the graviton from spinfoams :
 the 3D toy model}, J. high Energy Phys. JHEP05 (2006) 039, gr-qc/0512102; 
E. Livine and S. Speziale, {\em Group integral techniques for the spinfoam graviton propagator}, gr-qc/0608131

\bibitem{Freidel} Laurent Freidel and Etera R. Livine  {\em Ponzano-Regge model revisited III: Feynman diagrams and effective field theory}, Class. Quant. Grav. 23 2021-2062 (2006), hep-th/0502106 
  
\bibitem{FB} Aristide Baratin and Laurent Freidel, {\em Hidden quantum gravity in 4d Feynman diagrams: emergence of spin foams}, hep-th/0611042

\bibitem{giesel} K. Giesel and  T. Thiemann, 
{\em Algebraic quantum gravity (AQG) I. Conceptual setup}, gr-qc/0607099;
K. Giesel and T. Thiemann, {\em Algebraic quantum gravity (AQG) II. Semiclassical analysis}
gr-qc/0607100; K. Giesel and  T. Thiemann, 
{\em Algebraic quantum gravity (AQG) III. Semiclassical perturbation theory},
gr-qc/0607101

\bibitem{braids} Sundance O. Bilson-Thompson, Fotini Markopoulou, Lee Smolin,
{\it Quantum gravity and the standard model}, hep-th/0603022.  



\bibitem{Boj} M. Bojowald, {\em Inverse scale factor in isotropic quantum geometry}, Phys. Rev. D64 084018
(2001); M. Bojowald, {\em Loop Quantum Cosmology \textrm{IV}: discrete time evolution}, 
Class. Quant. Grav. 18, 1071 (2001); 
Martin Bojowald,  ``Loop quantum cosmology: recent progress", gr-qc/0402053

\bibitem{MAT} A. Ashtekar, M. Bojowald and J. Lewandowski, {\em Mathematica structure of loop quantum cosmology}, Adv. Theor. Math. Phys. 7 (2003) 233-268, gr-qc/0304074

\bibitem{work1} Leonardo Modesto, {\em Disappearance of the black hole singularity in loop  quantum gravity}, Phys. Rev. D 70 (2004) 124009, gr-qc/0407097

\bibitem{work2} Leonardo Modesto, {\em The kantowski-Sachs space-time in loop quantum gravity}, International Journal of Theoretical Physics, published on line 1 june 2006, 
 gr-qc/0411032
 
\bibitem{work3} Leonardo Modesto, {\em Gravitational collapse in loop quantum gravity}, 
gr-qc/0610074;
Leonardo Modesto, {\em Quantum gravitational collapse}, gr-qc/0504043

\bibitem{ABM} Leonardo Modesto, {\em Loop quantum black hole}, Class. Quant. Grav. 23 (2006) 5587-5602, gr-qc/0509078

\bibitem{BHI} Leonardo Modesto, {\em Black hole interior from loop quantum gravity}, 
gr-qc/ 0611043

\bibitem{ELQBH} Leonardo Modesto, {\em Evaporating loop quantum black hole}, 
gr-qc/0612084

\bibitem{variables} Abhay Ashtekar, {\em New Hamiltonian formulation of general relativity}, 
Phys. Rev. D 36 1587-1602



 \bibitem{LoopOld}  C. Rovelli and L. Smolin,
  {\em Loop Space Representation Of Quantum General Relativity}, 
  Nucl. Phys. B 331 (1990) 80; 
 C. Rovelli and L. Smolin,
{\em Discreteness of area and volume in quantum gravity}, 
Nucl. Phys. B 442 (1995) 593


\bibitem{AFW} A. Ashtekar, S. Fairhurst and J. Willis, {\em Quantum gravity, shadow states, and quantum mechanics}, Class. Quant. Grav. {\bf 20} 1031-1062 (2003)

\bibitem{Partial} C Rovelli, ``Partial observables", {Phy
Rev} {D65} (2002) 124013; gr-qc/0110035


\bibitem{Fonte.Math} H. Halvorson, {\em Complementary of representations in quantum mechanics}, Studies in History and Philosophy of Modern Physics 35, 45-56   (2004), quant - ph/0110102

\bibitem{KS} R. Kantowski and R. K. Sachs, J. Math. Phys. 7 (3) (1966)

\bibitem{Thie} T. Thiemann, {\em Quantum Spin Dynamics}, Class. Quant. Grav. {\bf 15}, 839 (1998) 


\bibitem{BojThiemann} I. Bengtsson,  ``Note on Ashtekar's variables in the spherically
symmetric case", Class. Quant. Grav. {\bf 5} (1988) L139-L142;
I. Bengtsson,  ``A new phase for general relativity?", Class. Quant. Grav. {\bf 7} (1990) 27-39;
H.A. Kastrup \& T. Thiemann,  ``Spherically symmetric gravity as a complete integrable system",
Nucl. Phys. B 425 (1994) 665-686, gr-qc/9401032;
M. Bojowald \& H.A. Kastrup,   ``Quantum symmetry reduction of Diffeomorphism invariant
theories of connections", JHEP 0002 (2000) 030, hep-th/9907041

\bibitem{BojImp}
M. Bojowald,  ``Spherically symmetric quantum geometry: states and basic operators",
Class. Quant. Grav. {\bf 21} (2004) 3733-3753, gr-qc/0407017


\bibitem{Bombelli} L. Bombelli \& R. J. Torrence,  ``Perfect fluids and Ashtekar variables,
with application to Kantowski-Sachs models", Class. Quant. Grav. {\bf 7} (1990) 1747-1745 



\end{thebibliography}
\end{document}